\documentclass[sigconf]{acmart}

\usepackage{subfigure}
\usepackage{graphicx}
\usepackage{textcomp}
\usepackage{xcolor}
\usepackage{tcolorbox}
\tcbuselibrary{skins}
\usepackage{float}
\usepackage{adjustbox}
\usepackage{etc}
\usepackage{pifont}

\usepackage{marginfix}

\usepackage[ruled,vlined]{algorithm2e}
\usepackage{algpseudocode}
\usepackage{float}
\usepackage{multirow}
\definecolor{lightergray}{RGB}{245,245,245}

\def\BibTeX{{\rm B\kern-.05em{\sc i\kern-.025em b}\kern-.08em
    T\kern-.1667em\lower.7ex\hbox{E}\kern-.125emX}}
    
\AtBeginDocument{%
  \providecommand\BibTeX{{%
    Bib\TeX}}}

\copyrightyear{2026}
\acmYear{2026}
\setcopyright{acmlicensed}
\acmConference[ICPC '26]{34th IEEE/ACM International Conference on Program Comprehension}{April 12--13, 2026}{Rio de Janeiro, Brazil}
\acmBooktitle{34th IEEE/ACM International Conference on Program Comprehension (ICPC '26), April 12--13, 2026, Rio de Janeiro, Brazil}
\acmPrice{}
\acmDOI{10.1145/3794763.3794818}
\acmISBN{979-8-4007-2482-4/2026/04}

\begin{document}

\title{Improved Bug Localization with AI Agents Leveraging Hypothesis and Dynamic Cognition}

\author{Asif Mohammed Samir}
\affiliation{%
	\institution{Dalhousie University}
	\city{Halifax}
	\state{NS}
	\country{Canada}
}
\email{asifsamir@dal.ca}

\author{Mohammad Masudur Rahman}
\affiliation{%
	\institution{Dalhousie University}
	\city{Halifax}
	\state{NS}
	\country{Canada}
}
\email{masud.rahman@dal.ca}

\begin{abstract}
Software bugs cost technology providers (e.g., AT\&T) billions annually and cause developers to spend roughly 50\% of their time on bug resolution. Traditional methods for bug localization often analyze the suspiciousness of code components (e.g., methods, documents) in isolation, overlooking their connections with other components in the codebase. Recent advances in Large Language Models (LLMs) and agentic AI techniques have shown strong potential for code understanding, but still lack causal reasoning during code exploration and struggle to manage growing context effectively, limiting their capability. In this paper, we present a novel agentic technique for bug localization --\textit{CogniGent}-- that overcomes the limitations above by leveraging multiple AI agents capable of causal reasoning, call-graph-based root cause analysis and context engineering. It emulates developer-inspired debugging practices (a.k.a., dynamic cognitive debugging) and conducts hypothesis testing to support bug localization. We evaluate CogniGent on a curated dataset of 591 bug reports using three widely adopted performance metrics and compare it against six established baselines from the literature. Experimental results show that our technique consistently outperformed existing traditional and LLM-based techniques, achieving MAP improvements of 23.33-38.57\% at the document and method levels. Similar gains were observed in MRR, with increases of 25.14-53.74\% at both granularity levels. Statistical significance tests also confirm the superiority of our technique. By addressing the reasoning, dependency, and context limitations, CogniGent advances the state of bug localization, bridging human-like cognition with agentic automation for improved performance.
\end{abstract}



\maketitle

\section{Introduction}
Software bugs pose major challenges in security and operational reliability, costing technology providers (e.g., AT\&T~\cite{atAndT_1,atAndT_2}, Microsoft~\cite{BI_crowdstrike}) billions of dollars every year. Experts from major technology companies (e.g., Google, Meta, Microsoft) identify bug resolution as one of the most pressing challenges in software development and maintenance~\cite{practitioners_bug_localization_study}. This concern is reinforced by industry studies showing that nearly 42\% of organizations lose $\approx$\$1 million annually due to poor software quality~\cite{tricentis2025quality}. This leads to developers spending about 50\% of their programming time on bug resolution~\cite{o2017debugging_stats, devops2024_dev_stats, britton2013reversible_stats}. Such debugging efforts are cognitively demanding, requiring reasoning and deep code understanding~\cite{cognitive_debugging_1, cognitive_debugging_2, cognitive_debugging_3_icse, cognitive_debugging_book}.

Software bugs are submitted to issue-tracking systems (e.g., Jira, Bugzilla, GitHub) as bug reports. These reports are the primary source of information for developers to understand the problem, inspect the project repository, and locate the buggy code. However, the quality of bug reports often varies depending on the reporter’s expertise in explaining the problem and writing style~\cite{blizzard}. As a result, even experienced developers relying on the bug reports may struggle to identify the faulty code~\cite{rahman2021forgotten}. A recent study with developers found that 79\% of debugging time was spent in long episodes (15--33 minutes) dedicated to locating and diagnosing bugs, but many of them ended without successfully identifying the faulty code~\cite{debugging_study}. Therefore, bug localization has been a long-standing challenge for developers. Unfortunately, automated approaches have not yet matured into reliable, practical tools to date~\cite{ciborowska_fbl_bert}.

Traditional methods for bug localization can be grouped into two main categories: spectrum-based and Information Retrieval (IR)-based approaches. Spectrum-based techniques rely on execution traces, which are not always available in practice~\cite{BRaIn, spectra2}. In contrast, IR-based approaches attempt to localize bugs by matching tokens in bug reports with those in the source code~\cite{ir_bug_localization1, ir_bug_localization3_topic, ir_bug_localization2_spectra, ir_localization_ml_dl}. While IR-based methods are lightweight and fast, they rely on surface-level token matching, which makes them prone to vocabulary mismatch between natural language and source code~\cite{furnas1987vocabulary}, especially due to the variability in bug reports. To address the limitations, several avenues have been explored, such as query reformulations~\cite{query_reformulation_tosem, sisman2012incorporating, rahman2021forgotten, chaparro2017_ob_eb}, incorporating statistical insights from code change history, and leveraging past bug fixes~\cite{bl_code_change, ir_ml_amalgam}. However, these techniques have not delivered significant improvements over earlier methods~\cite{bench4bl} and do not capture the deeper semantic context (e.g., program semantics~\cite{program_semantics}) needed for accurate localization of software bugs~\cite{BRaIn, dreamloc}.

Recent advancements in deep learning have shown great promise in understanding natural language and source code~\cite{code_generation, summarization}, which could support bug localization~\cite{ciborowska_fbl_bert, dreamloc}. However, training deep learning models requires large amounts of relevant, high-quality data~\cite{BRaIn}. In contrast, prompt-based and agentic AI techniques~\cite{sweBench, AutoCodeRover} present a promising alternative since they are powered by Large Language Models (LLMs), pre-trained on giant corpora. However, these approaches have been primarily optimized for complete issue resolution rather than detecting the location of bugs, which could lead to sub-optimal performance. Besides the above, contemporary approaches suffer from three major challenges as follows.

\textbf{(a) Determining fault-proneness of code components overlooking their dependencies:}
Traditional approaches for bug localization often determine the fault-proneness of a code component (e.g., method or source document) in isolation, overlooking its interactions with or dependencies on other components~\cite{BRaIn, ir_ml_amalgam, blizzard, ir_bug_localization1, ir_localization_lda_buglocator}. However, many bugs could trigger during runtime due to error propagation across components and could span across multiple components. Recent graph-based methods attempt to incorporate structural information to represent program flow and syntactic relationships~\cite{Grace_graph_bl, sgAttention_graph_bl}; however, their analysis often remains confined to the methods within the same source document, which might not be sufficient. Detection of such propagated bugs warrants an analysis of not only individual code components but also their dependencies within the codebase.

\textbf{(b) Lack of reasoning about root causes during code exploration:}
Bug reports often describe the failure symptoms of a software application rather than its faulty code since they are written by software users or testers~\cite{CoSIL, BRaIn}. Existing IR-based techniques rely on textual and semantic similarity between bug reports and source code documents to retrieve and detect faulty code~\cite{blaze, ir_bug_localization1}. Given their simplicity, these similarity measures might fail to capture the causal links between the reported symptoms and the faulty code~\cite{BRaIn}. Even recent agentic techniques (e.g., LocAgent~\cite{LocAgent}), powered by LLMs, assume the presence of code entities within the issue reports to choose the starting points for their code exploration. However, presence of such cues in bug reports cannot always be guaranteed~\cite{BRaIn}. Thus, such a lack of reasoning in existing techniques \cite{LocAgent} when finding candidates to explore could result in overlooking important dependencies or exploring irrelevant paths during bug localization~\cite{CoSIL}.

\textbf{(c) Lack of effective context management during code exploration:} During code navigation, agentic techniques perform iterative reasoning and extend the context length at each step with new information \cite{LocAgent, AutoCodeRover, CoSIL}. Although recent LLMs support much larger contexts (e.g., 128k tokens~\cite{llama3herd2024}), studies show that performance declines as context length increases~\cite{context_length_impact_1, context_rot_context_length_impact}. Irrelevant information can further worsen this effect~\cite{context_irrelevant_confuse_llm}. Thus, contemporary techniques might suffer from low performance and high inference costs during bug localization.

In this paper, we propose a novel technique, \textit{CogniGent}, that employs multiple AI agents and conducts hypothesis testing to localize software bugs. First, CogniGent emulates developers' debugging practices (a.k.a., dynamic cognitive debugging) and formulates multiple hypotheses on the \textit{root cause} of a bug based on the symptoms found in the bug report. Second, its AI agents apply causal reasoning to dynamically select starting points within the code and explore \textit{dependent code} (e.g., invoked methods) to assess their suspiciousness. To facilitate this process, our technique employs a novel algorithm —\textit{Click2Cause}— that traverses the call graph via depth-first exploration. It also applies \textit{scratchpad-based context management}~\cite{scratchpad_agent_langchain} to prevent context overload during code exploration. Finally, an independent observer agent evaluates the explored code chains, the generated artifacts or evidence, and determines their fitness to the hypotheses, delivering the final ranked list of faulty components at different granularity levels (e.g., methods, documents).

We evaluate our technique using 591 bug reports curated from an existing dataset of Samir et al.~\cite{samir2025improvingirbasedbuglocalization} and determine its performance at both the method and document levels using three widely adopted metrics: Mean Average Precision (MAP), Mean Reciprocal Rank (MRR), and HIT@K. We compared our approach with six baseline techniques~\cite{lucene,blizzard, saha_bleuir, BRaIn, Agentless, LocAgent} from two major areas--IR and LLM. CogniGent consistently outperformed existing traditional and LLM-based techniques, achieving MAP improvements of 23.33-38.57\% at the document and method levels. Similar gains were observed in MRR, with increases of 25.14-53.74\% across the same levels. These results underscore the effectiveness and superiority of CogniGent in software bug localization.

Thus, this research makes the following contributions-
\begin{itemize}
    \item A novel agentic emulation of \emph{dynamic cognitive debugging}, that represents human cognition during debugging by formulating and testing hypotheses and systematically reasoning about causal relationships between bugs and code.
    \item A novel agentic workflow, \textit{CogniGent}, that adopts dynamic cognitive debugging and employs Click2Cause, a repository-level code navigation algorithm, and scratchpad-based context management to effectively localize the faulty code (e.g., methods, documents).
    \item An extensive evaluation of CogniGent using three widely used metrics on a dataset of 591 recent bug reports annotated at both method and document levels.
    \item A replication package~\cite{cognigent_replication} containing the prototype, curated dataset, and configuration details for third-party use.
\end{itemize}

\begin{table}[ht]
  \centering
  \vspace{-0.5em}
  \caption{Bug Report \textit{(Apache HBase 15801258)}}
  \vspace{-0.5em}
  \label{tab:bug_report}

  {\fontsize{8}{9}\selectfont
  \resizebox{0.9\columnwidth}{!}{%
    \begin{tabular}{|p{8cm}|}
      \hline
      \rule{0pt}{2.5ex}\textbf{Title:} The failsafe snapshot should be deleted after rollback successfully.\\[2pt]
      \textbf{Description:} When a table exists and is in a disabled state, HBase supports restoring from a snapshot. In this scenario, HBase creates a failsafe snapshot for the disabled table and rolls back using it when restore fails. However, the failsafe snapshot remains even after successful rollback, and it should be deleted.\rule[-1.5ex]{0pt}{0pt}\\
      \hline
    \end{tabular}
  }}\\[0.3em]

  {\fontsize{6}{8}\selectfont
  \resizebox{0.9\columnwidth}{!}{%
    \begin{tabular}{|p{2.2cm}|p{3.8cm}|p{0.8cm}|}
      \hline
      \textbf{Technique} & \textbf{Approach} & \textbf{Rank} \\ \hline
      Baseline IR (Lucene \cite{lucene}) & Textual matching (Bug Report) & 87 \\ \hline
      Agentless \cite{Agentless} & Project structure cue + Semantic similarity & 35 \\ \hline
      LocAgent \cite{LocAgent} & Agentic & 8 \\ \hline
      \textbf{CogniGent} & Dynamic Cognitive Debugging & 1 \\ \hline
    \end{tabular}
  }}\\[0.7em]

  \includegraphics[width=0.9\columnwidth]{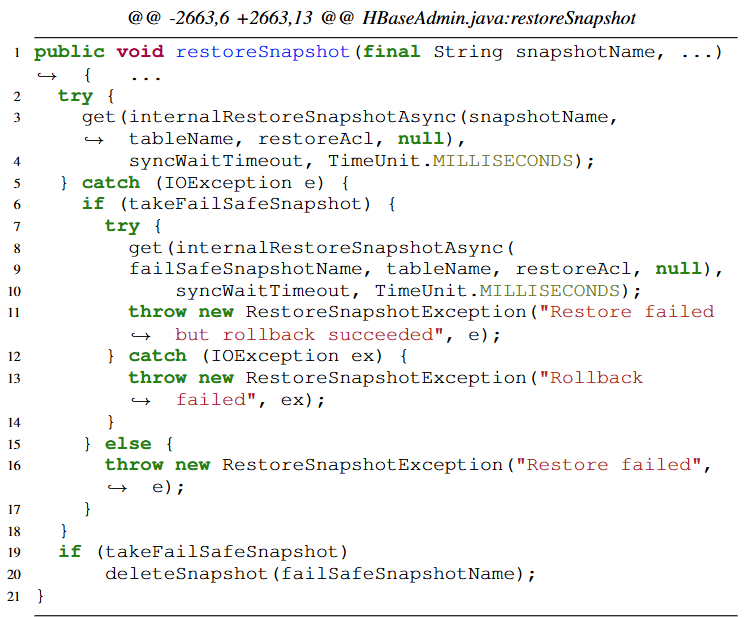}
  \captionof{figure}{Buggy Code for Bug \#15801258 from HBase}
  \label{fig:buggy_code}
  \vspace{-1.5em}
\end{table}

\section{Motivating Example}
In this section, we present a motivating example to demonstrate the effectiveness of our technique in bug localization. Let us consider a real bug report from the \textit{Apache HBase} project (check Table \ref{tab:bug_report}) that provides only a generic explanation of the bug. Here, the bug caused the failsafe snapshot of a table to survive from a rollback operation, leaving redundant artifacts in the cluster.

If we examine the source code in Fig.~\ref{fig:buggy_code}, at first glance, it appears correct--it calls \texttt{deleteSnapshot()} method after restoration, suggesting proper cleanup operation. However, given the reported bug, a developer might assume that the deletion may not be taking place after rollback. To verify this, they might navigate (e.g., \textit{Ctrl+Click}) to the \texttt{deleteSnapshot()} method  (line 20) for a close inspection, only to find it syntactically and logically correct. Given the role of the rollback operation, the developer might then inspect \texttt{internalRestoreSnapshotAsync()} method (line 3) to check whether it handles cleanup internally. As shown in Fig. 2, the method only initiates the restore or rollback operation but contains no deletion logic. The code contains two rollback operations within two-level nested try-catch blocks, complicating the control flows. When the first restore operation fails, an exception is caught, and a second attempt at restoration is made. When the second attempt at restoration is successful, a new \texttt{RestoreSnapshotException} is rethrown unexpectedly (line 13). This rethrow causes control flow to exit the method before reaching the \texttt{deleteSnapshot()}  (line 20) placed after the \texttt{try–catch} block. Even if \texttt{takeFailSafeSnapshot} is set to \texttt{true}, the deletion will not take place, and the failsafe snapshot will remain intact. 

This example highlights the importance of human cognitive processes in debugging, which involves formulating a hypothesis, collecting evidence, and testing the hypothesis.
Emulating these cognitive abilities through AI agents, our technique \textit{CogniGent} ranks the buggy \texttt{restoreSnapshot()} method first. Since the bug report lacks direct references to code elements, approaches relying on textual, semantic, or structural cues struggle to establish relevance. As a result, traditional IR-based baseline \cite{lucene} that relies solely on textual matching ranks it 87\textsuperscript{th}. Agentless~\cite{Agentless}, which integrates semantic similarity with project-structure cues, performs better but still ranks it 35\textsuperscript{th}. LocAgent~\cite{LocAgent}, an agentic approach that explores code using syntactic cues, ranks the method eighth.

\begin{figure*}[t]
    \centering
    \includegraphics[width=\textwidth]{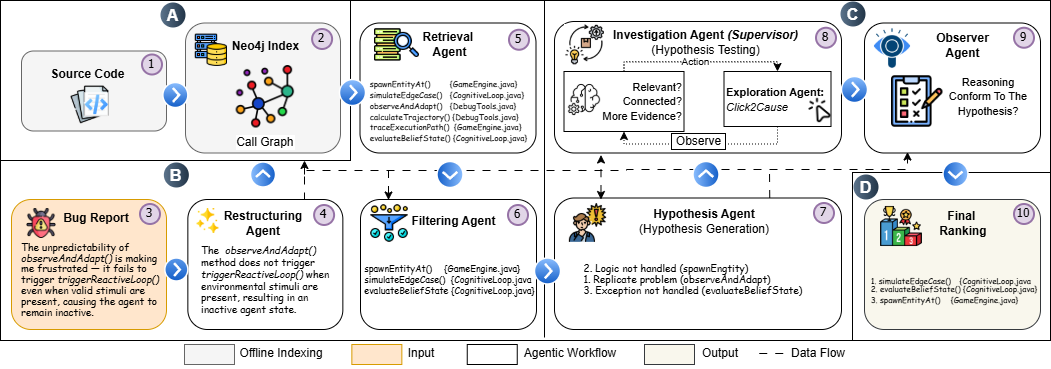}
    \caption{\centering 
        Schematic Diagram of \textit{CogniGent}:
(A) Graph-Based Indexing of Source Code,
(B) Retrieving and Filtering Documents,
(C) Dynamic Cognitive Debugging,
(D) Final Ranking.
    }
    \label{fig:schematic_diagram}
\end{figure*}

\section{Methodology}
Fig.~\ref{fig:schematic_diagram} shows the schematic diagram of our proposed technique, \textit{CogniGent}, for software bug localization. We discuss the major design steps of our technique as follows.

\subsection{Design of AI Agents}
\label{state_context}
In our agentic workflow, six role-specific agents form a pipeline to support bug localization: a restructuring agent (reorganizes a bug report), a retrieval agent (collects candidate code segments), a filtering agent (filters out irrelevant candidates), a hypothesis agent (generates hypotheses), an investigation agent (tests hypotheses), and an observer agent (validates hypotheses). We implement this workflow using the LangGraph library~\cite{langgraph2024}, combining varying sizes of LLMs (e.g., Devstral~\cite{devstral2024}) to balance efficiency, cost, and performance. The agents communicate through a shared LangGraph state~\cite{langgraph_state}, where each agent uses and updates the relevant parts of the state (e.g., adding hypotheses).

Since prompt design is central to the effectiveness of our workflow, we construct an appropriate system prompt~\cite{BRaIn} for each agent following the best practices of prompt-engineering ~\cite{prompt_best_practice, BRaIn} and employ few-shot chain-of-thought prompting~\cite{fewshot_CoT, CoT}. We apply meta-prompting~\cite{meta-prompting} to refine our prompts using LLMs (e.g., ChatGPT), as prompt quality can affect model performance~\cite{CoT, meta-prompting}.

\subsection{Graph-Based Indexing of Source Code} We collect candidate code segments (e.g., methods, constructors) of a software project using Information Retrieval (IR), which warrants the indexing of source code (Step 1-2, Fig.~\ref{fig:schematic_diagram}). Rather than treating the entire source code document as the retrieval unit~\cite{blizzard, BRaIn}, we follow Liu et al.~\cite{codeXgraph} and index code segments with their functionality and relationships. In particular, we use Neo4j~\cite{neo4j} for its graph construction capability and textual search, powered by the Lucene engine~\cite{lucene}. We construct the index as a directed heterogeneous graph \(G = (V, E)\), where nodes \(V\) represent code segments (methods, constructors), each uniquely identified by an ID, and edges \(E\) capture the connections among them. To extract these connections (e.g., invoke, inherit), we use JavaParser~\cite{javaparser}, which automatically identifies relationships through runtime type resolution. We indexed the source code of 132 versions from 15 software systems in the dataset to ensure that each bug was checked against the system and version for which it was originally reported~\cite{BRaIn}.

\subsection{Retrieving and Filtering Documents} 
\subsubsection{Restructuring Bug Reports} Several parts of bug reports (e.g.,  title, description) are often used as queries to retrieve potential buggy candidates. However, they might contain noisy or irrelevant information (e.g., emotional expressions, unrelated context), which can cause textual retrieval methods (e.g., Apache Lucene~\cite{lucene}) to return irrelevant code. Standard preprocessing techniques (e.g., stop-word removal, stemming) are insufficient for reducing such noise, as they risk losing the continuous textual understanding required by LLMs. Therefore, we employ a lightweight LLM (e.g., Qwen-Coder, 1.5B parameters) to restructure bug reports and remove their noise before the retrieval operation (Step 4, Fig.~\ref{fig:schematic_diagram}).

\subsubsection{Retrieving Potentially Buggy Segments} We use the restructured bug reports to retrieve the top (e.g., 100) potentially buggy code segments from our Neo4j index (Step 5, Fig.~\ref{fig:schematic_diagram}). We restrict the search to the system and version of the code associated with each bug report to avoid irrelevant results. We then pass the retrieved results to the subsequent filtering stage of our agentic workflow.

\subsubsection{Filtering Documents} 
We filter candidate segments using Intelligent Relevance Feedback (IRF)~\cite{BRaIn}, where a lightweight LLM reasons about the likelihood a code segment might be connected to the reported bug (Step 5, Fig.~\ref{fig:schematic_diagram}). Because bug reports often describe error symptoms~\cite{CoSIL} and keyword matching alone may miss related code~\cite{codeNL_semanticSearch_fail_1, codeNL_semanticSearch_fail_2}, we leverage the contextual reasoning ability of LLMs to identify faulty candidate segments~\cite{BRaIn}. This step produces a small set of top-ranked candidates (e.g., 10). 
\begin{table}[t]
\vspace{-0.3cm}
\centering
\footnotesize
\caption{Tools Used in CogniGent\vspace{-.4cm}}
\label{tab:tools}
\begin{tabular}{p{2cm}|p{1.2cm}|p{4.5cm}}
\hline
Tool Name & Used By & Function \\ \hline \hline
\textit{retrieveCandidates} & Retrieval Agent & Retrieves the potential buggy source code segments (e.g., methods) from the bug reports. \\ \hline
\textit{exploreCallChain} & Supervisor Agent & Determines and returns the most relevant chain of code segment(s). \\ \hline
\textit{getCodeSegment} & Exploration Agent & Retrieves the source code of a target code segment located in the neighborhood of the current segment under exploration. \\ \hline
\end{tabular}
\vspace{-0.5cm}
\end{table}

\subsection{Dynamic Cognitive Debugging} In our proposed technique, we apply Dynamic Cognitive Debugging (DCD) to bug localization. Prior studies identify debugging as a cognitively demanding task that requires programmers to construct and refine mental models to locate and fix errors~\cite{cognitive_debugging_1, cognitive_debugging_2, cognitive_debugging_3_icse, cognitive_debugging_book}. This process involves forming hypotheses, navigating the codebase to test them, and confirming or discarding those hypotheses based on relevant evidence. To emulate this important cognitive process, we leverage the reasoning capability of LLMs as AI agents and employ them for code navigation and bug detection as follows.

\subsubsection{Hypothesis Generation} During bug resolution, human experts typically form multiple hypotheses about the possible causes of a bug based on the bug report, available code, and then investigate further~\cite{cognitive_debugging_1}. Similarly, we provide the bug report and the filtered code segments above (Step 6) to a powerful LLM (e.g., Devstral 20B parameters), and leverage the model to generate multiple \textit{competitive hypotheses} (Step 7, Fig.~\ref{fig:schematic_diagram}). For each code segment, we guide the LLM to generate a hypothesis on the root cause of the reported bug, along with a confidence category (e.g., high, medium, or low) and a confidence score (e.g., 0-1). These categories represent the relative likelihood of each segment being buggy compared to the other candidates. We then retain the segments with high or medium confidence for deeper investigation in the subsequent steps.

\subsubsection{Investigation: Hypothesis Testing using AI Agents} In this step, we simulate how developers investigate and navigate code during debugging (Step~8, Fig.~\ref{fig:schematic_diagram}). For example, when a bug report mentions a \textit{race condition}, a developer may reason about it, inspect code related to threading functionality and then follow the relevant call chain for closer examination. Similarly, our technique applies contextual reasoning and autonomously determines whether the current segment provides sufficient evidence or whether parts of it (e.g., invoked methods) should be explored further.

We model this cognitive process with two specialized agents: \textit{a supervisor} and \textit{an explorer}. The supervisor is built on the ReAct framework (think, act, observe)~\cite{ReAct} to guide the exploration agent. For each retained code segment from the previous step (Step 7, Fig. \ref{fig:schematic_diagram}), a dedicated \textit{supervisor} agent is instantiated with its associated hypothesis and the bug report as part of its reasoning context. The agent reviews the segment, collects evidence, and determines whether further exploration is required to test the hypothesis. When an additional analysis is necessary, the agent first identifies the calls to follow from the current code segment, reasoning about their relevance to the reported symptoms. It then dynamically instantiates an independent \textit{exploration agent} and delegates the exploration task. The agent receives the bug report, the current code segment, the selected calls to explore prioritized by suspiciousness, and a dynamically determined maximum depth. We implement the exploration agent as a tool (Table \ref{tab:tools}) that maintains a separate scratchpad-based \cite{scratchpad_agent_langchain, ReAct} reasoning context focused exclusively on relevant code segments, thereby preventing context confusion.~\cite{context_confusion_1}. The scratchpad is separate from CogniGent’s pipeline state (see ~\ref{state_context}) and exists only for the lifetime of the exploration agent, maintaining its reasoning traces during exploration.

To gather more information targeting a hypothesis, the exploration agent uses our proposed Click2Cause algorithm (Alg.~\ref{alg:click2cause}) to traverse the call graph recursively. Unlike LocAgent~\cite{LocAgent}, which employs a breadth-first search (BFS) strategy~\cite{cormen_algorithms}, Click2Cause applies a depth-first search (DFS)~\cite{cormen_algorithms}, mimicking the \textit{Ctrl+Click} navigation familiar to developers \cite{ctrl_click}. The algorithm navigates method-call relationships stored in Neo4j, following a recursive \textit{think–act cycle}. It hops through the call graph by making tool calls (Table~\ref{tab:tools}) that retrieve the next callee method and its source code for inspection, and controls the maximum exploration depth along any single branch based on LLM's reasoning. During traversal, the exploration LLM assigns confidence scores to each segment-chain based on its alignment with the reported bug symptoms. When a branch becomes less promising (i.e., chain confidence drops), Click2Cause backtracks to the previous decision point, prunes the current path from the scratchpad, and continues exploring alternative sibling branches. This backtracking and pruning process ensures that only promising paths remain in the agent’s working memory. Conversely, when a confidence score exceeds a threshold specified in the prompt (e.g., 90\%), the exploration agent terminates early and returns the identified call chain to the supervisor agent. Otherwise, it returns the most promising chain observed.

Upon receiving a call chain, the supervisor agent reasons over it with respect to the hypothesis and the reported symptoms, and assigns a score. If the evidence supports the hypothesis, the chain is accepted as the likely root cause; otherwise, the supervisor either initiates further exploration or concludes based on the best evidence currently available.

This coordinated delegation strategy enables autonomous investigation to capture the buggy call chain through systematic context engineering~\cite{context_engineering}. In contrast to contemporary techniques~\cite{LocAgent}, our technique uses causal reasoning from the reported symptoms to determine candidates to explore.

\subsubsection{Observer} 
Once all investigations are complete, an independent observer agent evaluates each candidate explanation--either a single code segment or a call chain—along with its associated hypothesis and reasoning traces (Step~8, Fig.~\ref{fig:schematic_diagram}). The observer assesses how well each hypothesis is supported by the corresponding code and reasoning, and assigns a confidence score. The final suspiciousness score combines the supervisor’s reasoning confidence with the observer’s validation score, mitigating bias from either stage to find the most plausible root cause.

\subsection{Final Ranking} Based on the confidence scores assigned by the observer agent, we rank the code segments and their associated call chains while preserving the order of related segments within each chain. Since investigation is performed on a limited set of segments filtered through hypothesis generation, and because our evaluation focuses on the top-K methods (e.g., top-10), any remaining positions are filled with other candidates in descending order of confidence, first from the hypothesis generation stage (Step 7, Fig.~\ref{fig:schematic_diagram}) and then from the filtering stage (Step 6, Fig.~\ref{fig:schematic_diagram}), while avoiding redundancy. Finally, to obtain document-level rankings, we map these methods to their corresponding documents in the same confidence order to determine the top-K documents. This process produces the final ranked list of locations most likely to be buggy at both the method and document levels.

\begin{algorithm}[t]
\LinesNumbered
\caption{Click2Cause: Call Chain Analysis}
\label{alg:click2cause}
\small
\SetKwFunction{FMain}{Click2Cause}
\KwIn{$bugReport$, $startSeg$, $callsToExplore$, $maxDepth$, confidence threshold $\tau$, initial confidence $\mathcal{C}_{parent}$}
\KwOut{Most probable call chain $C^*$ with confidence $\mathcal{C}_{LLM}(C^*)$}
\textbf{Global:} 
$visited \gets \emptyset$, 
$scratchPad \gets \emptyset$, 
$C^* \gets (\emptyset, 0)$\\
\ForEach{$call \in callsToExplore$}{
    \textbf{DFS}$(startSeg, [call], depth{=}1, \mathcal{C}_{parent}, bugReport)$
}
\Return{$C^*$} \tcp{Return best call chain}
\BlankLine
\textbf{Recursive DFS$(seg, path, depth, \mathcal{C}_{parent}, bugReport)$:}\\
\Indp
\lIf{$seg \in visited$ \textbf{or} $depth > maxDepth$}{\Return}
Add $seg$ to $scratchPad$, $visited$, and $path$\\
$\mathcal{A}(seg) \gets \text{LLMReason}(bugReport, path)$\\
$\mathcal{C}_{LLM}(seg) \gets \mathcal{A}(seg).\text{conf}$\\
\eIf{$\mathcal{C}_{LLM}(seg) < \mathcal{C}_{parent}$}{
    \textbf{Backtrack()} \tcp{Prune weak branch}
}{
    \If{$\mathcal{C}_{LLM}(seg) > \mathcal{C}_{LLM}(C^*)$}{
        $C^* \gets (path, \mathcal{C}_{LLM}(seg))$
    }
    \If{$\mathcal{C}_{LLM}(seg) \ge \tau$}{\Return \tcp{early stop}}
    \ForEach{$next \in \mathcal{A}(seg).\text{callsToExplore}$}{
        \textbf{DFS}$(next, path, depth{+}1, \mathcal{C}_{LLM}(seg), bugReport)$
    }
}
\Return\\
\Indm
\end{algorithm}

\begin{table}[ht]
    \caption{Dataset Summary\vspace{-.3cm}}
    \centering
    \begin{tabular}{|c|c|c|c|c|c|c|}
        \hline
        Project & Systems & Versions & Count & NL & PE & ST \\
        \hline \hline
        Apache   & 4  & 17  & 104 & 60  & 23  & 21 \\
        Spring   & 7  & 92  & 422 & 206 & 163 & 53 \\
        Wildfly  & 4  & 23  & 65  & 33  & 14  & 18 \\
        \hline \hline
        Total    & 15 & 132 & 591 & 299 & 200 & 92 \\
        \hline
    \end{tabular}
    
    \vspace{0.5em}
    {\footnotesize \textit{NL = Natural Language, PE = Program Elements, ST = Stack Traces}}
    \label{tab:dataset_stats}
    \vspace{-0.5cm}
\end{table}

%

\section{Experiments}
We curated a dataset of 591 bug reports from the work of Samir et al.~\cite{samir2025improvingirbasedbuglocalization}, who extended the Bench4BL dataset \cite{bench4bl} with recent reports. For evaluation, we employ three widely used metrics from the literature: Mean Average Precision (MAP), Mean Reciprocal Rank (MRR), and HIT@K (K=1, 5, 10). We experiment with three different LLMs and compare our solution, CogniGent, against six relevant baseline techniques to demonstrate its effectiveness. Through these experiments, we address three research questions:

\begin{itemize}
\item \textbf{\textit{RQ$_1$}}: \textit{(a)} How does CogniGent perform in localizing software bugs? \textit{(b)} How does it perform on different types of bugs? \textit{(c)} How effective is it in localizing bugs that span multiple documents?
\item \textbf{\textit{RQ$_2$}}: How do the individual agentic modules of CogniGent contribute to its overall performance in bug localization?
\item \textbf{\textit{RQ$_3$}}: Can CogniGent outperform the relevant baseline techniques in bug localization?
\end{itemize}


\subsection{Dataset Construction}
Recent advances in agent-based solutions have motivated the construction of new datasets (e.g., SWE-bench~\cite{sweBench}). However, these datasets were primarily created for issue-fixing tasks and contain not only bug reports but also feature requests, questions, and other tasks~\cite{bug_vs_feature}. Such a mixture of issues can negatively affect the performance of bug localization techniques~\cite{bug_vs_feature}. Therefore, we curated a dataset composed exclusively of bug reports for our investigation.

We curated our dataset based on the extended version of Bench4BL constructed by Samir et al.~\cite{samir2025improvingirbasedbuglocalization}, which includes $\approx$1.7k bug reports from 2018–2024. These reports were collected from Jira and GitHub issue tracking systems. We first confirmed the existence of ground-truth documents (e.g., faulty code) corresponding to the bug reports. While the original dataset contained only ground-truth documents, we augmented it with the methods modified during bug resolution and separated test components (e.g., methods, documents) as distinct entries to enhance applicability. Since agentic systems are computationally demanding and indexing multiple software versions is challenging, we selected specific versions of the subject systems to enable version-based retrieval, while ensuring the dataset contains diverse bug report types (e.g., natural language descriptions and program elements)~\cite{blizzard}. We spent about 22 hours validating and refining the resulting dataset comprising 591 bug reports across 132 versions of 15 Java-based systems. Table \ref{tab:dataset_stats} provides the summary of our curated dataset.

\subsection{Evaluation Metrics}
\subsubsection*{\textbf{Mean Average Precision (MAP)}}

Precision@K measures the accuracy of retrieved results up to rank-K for each occurrence of a buggy source code component (e.g., method or document) in the ranked list. Average Precision is the mean of these Precision@K values across all relevant (buggy) items for a given query. Consequently, the Mean Average Precision (MAP) is obtained by averaging the Average Precision scores across all queries. 
Q in the dataset.

{\small
\vspace{-.5cm}
\begin{align*}
AP@K &= \frac{1}{|D|} \sum_{k=1}^{K} P_k \times B_k & \bigg| & \quad 
MAP &= \frac{1}{|Q|} \sum_{q=1}^{Q} AP@K_q
\end{align*}
}

Here, $AP@K$ is the average precision within the top-$K$ results, where $P_k$ is precision at rank $k$ and $B_k$ indicates if the $k^{th}$ item is buggy (1) or not (0). $MAP$ averages this across all queries $q$ in $Q$, with $D$ as the ground-truth set.

\subsubsection*{\textbf{Mean Reciprocal Rank (MRR)}}

Reciprocal Rank (RR) represents the position of the first relevant code component returned by a given method. It is calculated as the inverse of that component’s rank within the ordered list of retrieved items for each query.

\vspace{-0.5cm}
{\small
\begin{align*}
RR_q &= \frac{1}{\textit{Rank of First Relevant Item}} & \bigg| & \quad 
MRR = \frac{1}{|Q|} \sum_{q=1}^{|Q|} RR_q
\end{align*}
}

Here, $MRR$ is the mean of all Reciprocal Ranks ($RR_q$) computed across the set of queries $Q$.

\subsubsection*{\textbf{HIT@K}}

HIT@K~\cite{saha_bleuir} quantifies the percentage of queries for which at least one relevant component appears within the top-\( K \) retrieved results. Higher HIT@K values indicate better performance in bug localization techniques.

\vspace{-0.4cm}
{\small
\begin{align*}
HIT@K &= \frac{1}{|\mathcal{Q}|} \sum_{q=1}^{|Q|} \begin{cases} 
1, & r_q \in \mathcal{G} \\ 
0, & \text{otherwise} 
\end{cases} 
\end{align*}
}

Here, $r_q$ equals 1 if a query $q$ retrieves a ground truth item in the top-$K$ results (0 otherwise), in the set of all queries $Q$.


\subsection{Experimental Setup}
For our experiments, we selected three open-weight, instruction-tuned LLMs of varying scales with agent tool-calling support: LLaMA 3.3 70B~\cite{llama3herd2024}, a general-purpose model; Devstral 20B~\cite{devstral2024}, a Mistral variant fine-tuned for software engineering tasks; and Qwen3-Coder 30B~\cite{qwen3coder2024}, optimized for coding-related reasoning. These models have different types of specialization to support the interpretation of bug reports and the understanding of source code to aid bug localization, while also considering computational efficiency.

To balance performance and resource utilization, we adopted a hybrid execution setup. We used a lightweight 1.5B-parameter Qwen2.5-coder model on a local GPU-enabled system (NVIDIA GeForce RTX 2060) to perform restructuring and filtering tasks, while the larger models mentioned above were executed via a cloud-based multi-model gateway to handle reasoning and localization. The generation temperature was fixed at 0.5 to balance factual accuracy and creative reasoning~\cite{temperature_effect}  throughout all model interactions.

\subsection{Evaluating CogniGent}

\begin{table*}[t]
    \caption{Performance and Efficiency of CogniGent at Different Levels of Granularity\vspace{-.3cm}}
    \centering
    \begin{tabular}{|c|c|c|c|c|c|c|c|c|}
        \hline
        \textbf{Techniques} & \textbf{Level} & \textbf{MAP} & \textbf{MRR} & \textbf{HIT@1} & \textbf{HIT@5} & \textbf{HIT@10} & \textbf{Cost/Instance ($\bar{\$}$)} & \textbf{Time/Instance ($\bar{t}$)} \\
        \hline \hline
        \multirow{2}{*}{Baseline-IR (Lucene)}         
            & Document   & 0.330 & 0.334 & 0.254 & 0.443 & 0.487 & \multirow{2}{*}{--} & \multirow{2}{*}{--} \\
            & Method & 0.163 & 0.165 & 0.122 & 0.217 & 0.246 &  &  \\
        \hline
        \hline
        \multirow{2}{*}{CogniGent (LLaMA)}    
            & Document   & 0.345 & 0.356 & 0.283 & 0.455 & 0.509 & \multirow{2}{*}{0.0011} & \multirow{2}{*}{2m 43s} \\
            & Method & 0.165 & 0.185 & 0.135 & 0.232 & 0.268 &  &  \\
        \hline
        \multirow{2}{*}{CogniGent (Qwen-Coder)} 
            & Document   & 0.400 & 0.412 & 0.360 & 0.470 & 0.515 & \multirow{2}{*}{0.0058} & \multirow{2}{*}{3m 50s} \\
            & Method & 0.222 & 0.244 & 0.212 & 0.271 & 0.285 &  &  \\
        \hline
        \multirow{2}{*}{CogniGent (Devstral)} 
            & Document   & 0.407 & 0.418 & 0.374 & 0.475 & 0.526 & \multirow{2}{*}{0.0026} & \multirow{2}{*}{2m 58s} \\
            & Method & 0.226 & 0.254 & 0.227 & 0.272 & 0.289 &  &  \\
        \hline
    \end{tabular}
    \label{tab:performance_RQ1}
    \vspace{-.3cm}
\end{table*}

\subsubsection*{\textbf{Answering RQ$_1$ - Performance of CogniGent}} \label{sec_RQ1} We evaluate the performance of our proposed technique, CogniGent, with three key metrics- Mean Average Precision (MAP), Mean Reciprocal Rank (MRR), and top-K results (K = 1, 5, 10). Table \ref{tab:performance_RQ1} summarizes the performance of our technique using three different LLMs.

From Table~\ref{tab:performance_RQ1}, we observe that CogniGent performs effectively in localizing bugs across different levels of granularity. Among all variants, CogniGent (Devstral) demonstrates the strongest performance, achieving a Mean Average Precision (MAP) of 0.407 at the document level. This reflects CogniGent’s ability to rank buggy source documents higher than irrelevant ones. The Mean Reciprocal Rank (MRR) of 0.418 further indicates that the first relevant document typically appears within the top two results. Moreover, its HIT@1 score of 0.374 shows that, for approximately 37.4\% of the bugs, the buggy document is ranked at the very top. The performance remains consistent when more results are considered, with HIT@5 = 0.475 and HIT@10 = 0.526, demonstrating CogniGent’s strong localization ability. The Qwen-Coder variant exhibits comparable results, trailing slightly behind Devstral variant by 0.19\%–1.72\% across all metrics. On the other hand, CogniGent (LLaMA) performs moderately well while outperforming the traditional IR-based baseline, achieving around 15\% improvement in MAP and MRR, and up to 11.4\% improvement in HIT@1. However, it still lags behind the Devstral and Qwen-Coder variants. At the method level, CogniGent (Devstral) also demonstrates substantial gains, improving over the IR baseline by 1.23\%–53.94\% across all metrics. Similar trends are observed for Qwen-Coder, with improvements ranging from 12.12\% to 53.93\%, indicating strong robustness even at finer code granularity.

In terms of efficiency, CogniGent (LLaMA) achieves the lowest cost per instance (i.e., bug report) at \$0.0011 (0.11 cents) with an average runtime of 2 minutes 43 seconds (Table \ref{tab:performance_RQ1}). In comparison, Devstral costs 0.26 cents per instance with a comparable runtime, while Qwen-Coder, despite its similar performance, incurs more than double the cost and requires approximately 29\% more time for execution. During our experiments, we observed that the LLaMA variant makes fewer tool calls, which possibly explains their lower cost, while their general-purpose design may contribute to their lower performance.

\begin{figure}[t]
  \centering
  \begin{tabular}{c}   
    \hspace{-1em}
    \vspace{-0.5em}
    \includegraphics[width=0.44\textwidth]{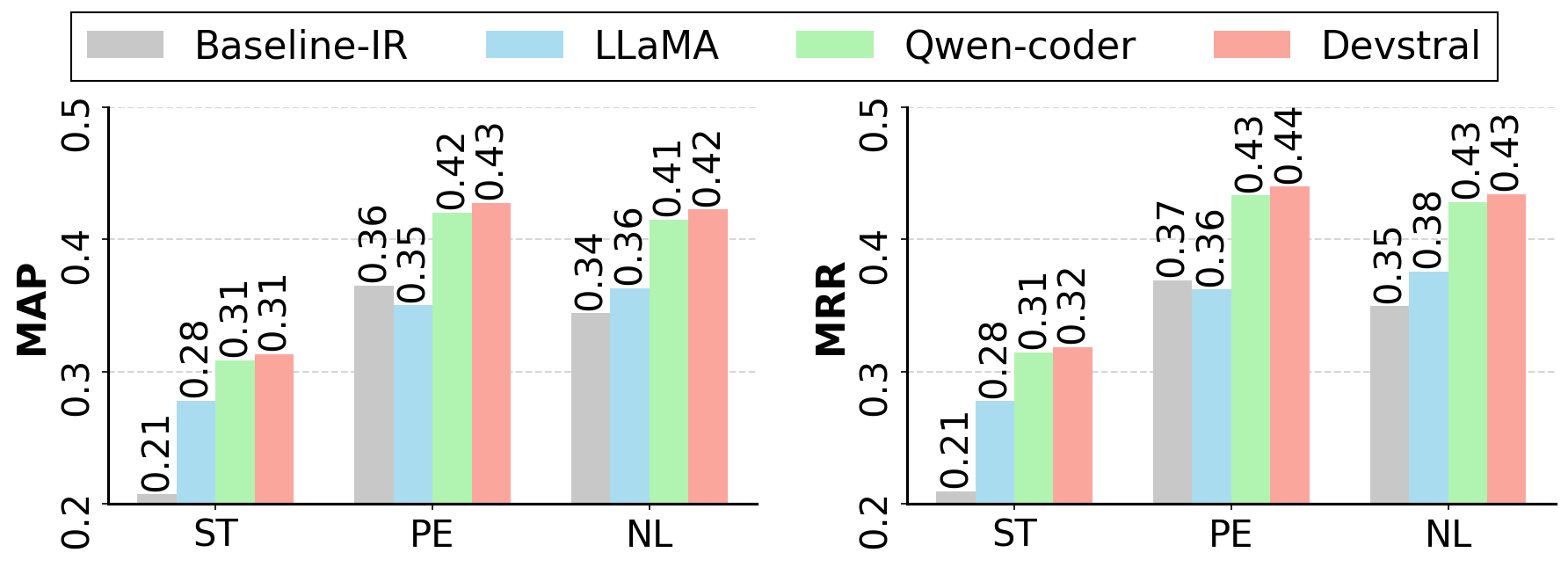} \\[2pt]
    
    \textbf{(a)}~Document-level performance \\[6pt]
    \hspace{-1em}
    \vspace{-0.5em}
    \includegraphics[width=0.44\textwidth]{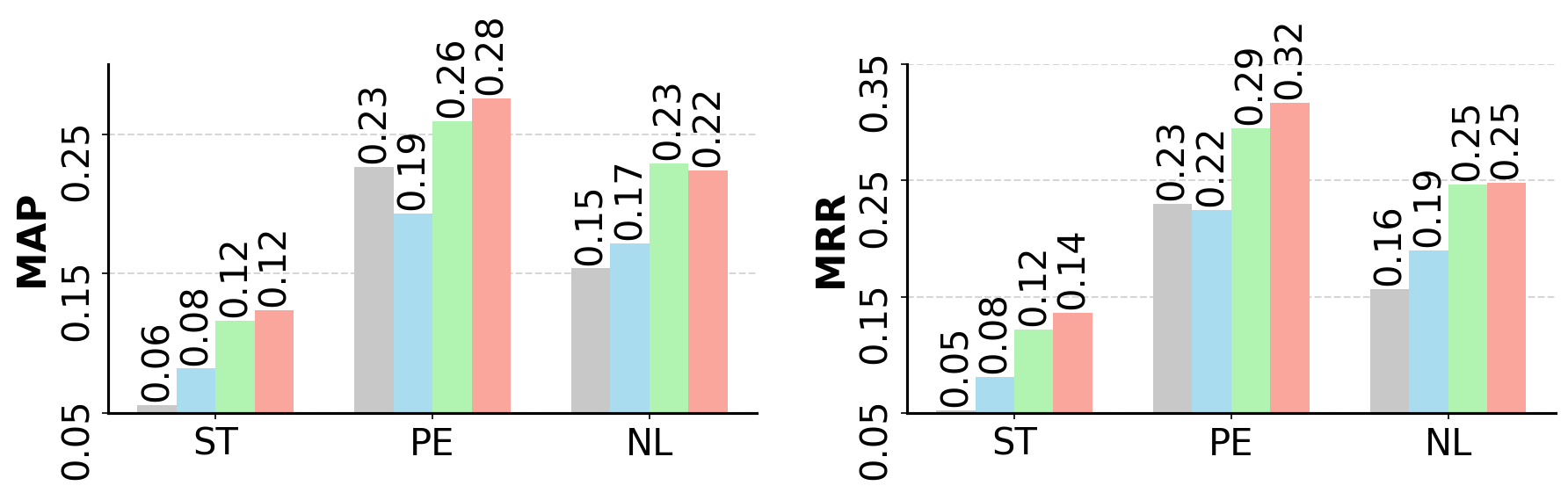} \\[2pt]
    \textbf{(b)}~Method-level performance \\
  \end{tabular}
  \vspace{-0.2cm}
  \caption{\centering CogniGent's Localization Performance Across Three Bug Types }
  \label{tab:bug_type_comparison_RQ1}
  \vspace{-0.3cm}
\end{figure}

Existing work has categorized bug reports into three categories based on the presence of structured entities in them -- Stack Trace (ST), Program Elements (PE), and Natural Language (NL) types~\cite{blizzard}. We categorize the reports using regular expressions (regex)~\cite{blizzard}. To demonstrate the robustness of our technique, we also evaluated its performance across these bug types.
Fig.~\ref{tab:bug_type_comparison_RQ1} shows CogniGent’s bug localization performance in terms of MAP and MRR. Our technique exhibits consistent trends across all variants, performing best on PE and NL reports while struggling with ST reports at the document level. For ST-type reports, CogniGent (Devstral) achieves a MAP of 0.31--about 34\% higher than baseline IR and up to 14.28\% higher than the other two variants. A similar pattern is observed in MRR, confirming that bug reports with stack traces remain the most difficult to localize.
In contrast, the Devstral and Qwen-Coder variants report higher gains for PE and NL, performing comparably with MAP and MRR improvements of 17.15\% and 19.09\%, respectively. A similar trend appears at the method level, though overall scores decline due to finer granularity. CogniGent (Devstral) remains dominant across all types of bug reports, achieving MAP values of 0.28 for PE and 0.22 for NL, while the Qwen-Coder variant trails by 4--6\%. CogniGent (LLaMA) shows a 25--40\% drop compared to the other two variants but still improves over baseline IR by up to 0.34\% for ST and 0.32\% for NL, with a slight decrease for PE. Thus, despite the increased difficulty of method-level localization, CogniGent presents a better alternative than the baseline for bug localization.

\begin{figure}[!ht]
    \centering
    \begin{tabular}{cc}
        \multicolumn{2}{c}{
            \hspace*{-1cm}\includegraphics[width=1\linewidth]{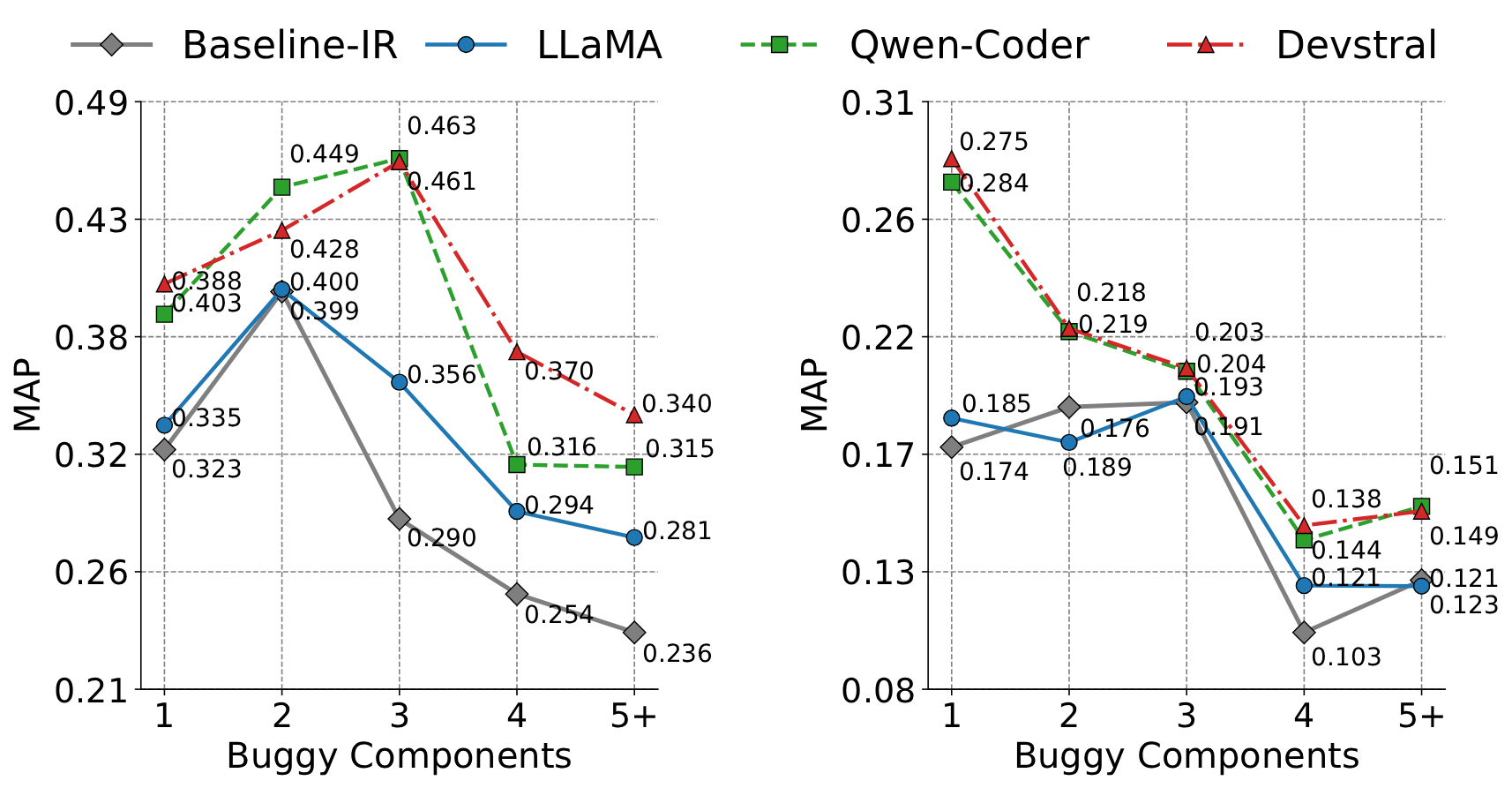}
        } \\
        (a) Document-level  & (b) Method-level 
    \end{tabular}
 
    \caption{\centering Performance for Single vs. Multi-Component Bugs}
    \label{fig:multi_Document_performance_RQ1}
    \vspace{-0.5cm}
\end{figure}

Given the prevalence of propagated bugs, we examined CogniGent's performance on bugs spanning multiple source components (e.g., methods, documents). We first grouped bug reports by their ground-truth counts-1, 2, 3, 4, or 5+ buggy documents or segments (e.g., methods) --and evaluated our techniques' performance using Mean Average Precision (MAP) at both document and method levels.
Fig.~\ref{fig:multi_Document_performance_RQ1} shows CogniGent’s performance across these groups of bug reports. CogniGent (Devstral) achieves a MAP of 0.40 for single-document bugs, improving to 0.43 and 0.46 for two and three documents (6.45\% and 9.42\% gains), but drops to 0.37 and 0.34 for four and five or more. Across all cases, it outperforms the IR baseline by 7.36–58.99\%. Qwen-Coder follows a similar trend—starting at 0.39, surpassing Devstral by 4.6\% for two-document bugs—and reaches up to 59.51\% improvement over IR. LLaMA variant reports lower MAP scores overall but still achieves 0.28–19\% gains over IR.
At the method level, scores decline as bugs span more segments. CogniGent (Devstral) remains best, starting at 0.28 for single-segment bugs and dropping to 0.14–0.15 for broader scopes, with up to 39.70\% improvement over IR. Qwen-Coder trails by 2--4\%, while LLaMA drops to 0.12 for large bugs yet maintains 1.87–6.38\% gains in other cases. Overall, while bug spread increases the difficulty of method-level reasoning, our technique effectively handles such cases through inter-component exploration and reasoning, achieving consistently strong localization performance.

\begin{table}[t]
	\centering
	\caption{Effects of Individual Components in CogniGent\vspace{-.2cm}}
	\label{tab:ablation_study_RQ2}
	\resizebox{\columnwidth}{!}{%
		\begin{tabular}{|c|c|c|c|c||c|c|}
			\hline
			\textbf{Restru-} & \textbf{Filt-} & \textbf{Hypo-} & \textbf{Investi-} & \textbf{Obser-} & \multicolumn{2}{c|}{\textbf{MAP}} \\
			\textbf{cturing} & \textbf{ering} & \textbf{thesis} & \textbf{gation} & \textbf{vation} & \textbf{Document} & \textbf{Method} \\
			\hline \hline
			\ding{51} & \ding{51} & \ding{51} & \ding{51} & \ding{51} & \textbf{0.407} & \textbf{0.226} \\
			\hline \hline
			-- & \ding{51} & \ding{51} & \ding{51} & \ding{51} & 0.382 & 0.212 \\
			\hline
			\ding{51} & -- & \ding{51} & \ding{51} & \ding{51} & 0.364 & 0.189 \\
			\hline
			\ding{51} & \ding{51} & -- & \ding{51} & \ding{51} & 0.379 & 0.195 \\
			\hline
			\ding{51} & \ding{51} & \ding{51} & -- & \ding{51} & 0.360 & 0.183 \\
			\hline
			\ding{51} & \ding{51} & \ding{51} & \ding{51} & -- & 0.392 & 0.216 \\
			\hline
		\end{tabular}%
	}
	\vspace{-0.3cm}
\end{table}  

 \begin{tcolorbox}[colback=lightergray,colframe=black,arc=1mm,boxrule=0pt,
     leftrule=2pt, bottomrule=1pt]
 \textbf{RQ1 Summary:} 
 CogniGent significantly improves bug localization at various granularity levels, particularly with Devstral, reaching a high MAP score of 0.407. This performance can be attributed to our technique’s ability to reason across multiple source code segments (e.g., methods) while dynamically exploring inter-component dependencies to discover root causes of failures for bug localization.
 \end{tcolorbox}
 \vspace{-0.2cm}

\subsubsection*{\textbf{Answering RQ$_2$ - Contribution of Different Modules}} CogniGent consists of five major agentic modules that collectively contribute to bug localization.
In this section, we evaluate the contribution of each module through an ablation study.
Considering the trade-off among cost, time, and performance, we report the results only for CogniGent (Devstral). Table \ref{tab:ablation_study_RQ2} reports the contribution of each module towards our technique's performance.

From Table~\ref{tab:ablation_study_RQ2}, we observe that all modules contribute positively to the localization precision (a.k.a., MAP) of CogniGent, though to varying degrees.
At the document level, removing the investigation module (Step 8, Fig. \ref{fig:schematic_diagram}) causes the largest drop in MAP to 0.360, which is an 11.57\% decrease. Removing the filtering or hypothesis generation modules (Step 6 and 7, Fig. \ref{fig:schematic_diagram}) also leads to moderate drops, reducing MAP to 0.364 and 0.379, respectively. This shows that both modules are important for selecting the right candidates and ranking them effectively. In comparison, the absence of the restructuring and observer modules (Step 4 and 9, Fig. \ref{fig:schematic_diagram}) reduces CogniGent’s MAP scores to 0.382 and 0.392 (6.14\% and 3.69\% drops), indicating that although these modules contribute to the overall performance, their effects are comparatively smaller.

We see a similar pattern at the method level. Removing the investigation module causes the largest decrease in MAP to 0.183, a 19.02\% drop. The filtering and hypothesis modules also remain important, reducing MAP by 16.37\% and 13.71\% to 0.189 and 0.195, respectively. The restructuring and observer modules make smaller but consistent contributions, with MAP values of 0.212 and 0.216.

Overall, these findings show that the hypothesis generation and investigation modules (Steps 7–8, Fig.~\ref{fig:schematic_diagram}), core parts of dynamic cognitive debugging, contribute substantially to bug localization by reasoning about hypotheses and gathering evidence across connected code segments. However, the ablation results also highlight that the overall effectiveness of CogniGent relies on the collective contribution of all modules working together.

 \begin{tcolorbox}[colback=lightergray,colframe=black,arc=1mm,boxrule=0pt,
     leftrule=2pt, bottomrule=1pt]
 \textbf{RQ2 Summary:} CogniGent’s ablation study shows that hypothesis generation and hypothesis testing drive the largest gains (up to 19\%). Smaller modules, such as restructuring and observation, still offer steady improvements, indicating that CogniGent’s effectiveness arises from the interplay among its modules.
 \end{tcolorbox}

\subsubsection*{\textbf{Answering RQ$_3$ - Comparison with Baseline Technique}} To position our work within the broader research landscape, we compare our technique with relevant baseline methods from prior studies. Table~\ref{tab:RQ3_baseline_comparison} compares the performance in terms of MAP, MRR, and HIT@K (K=1, 5) of our technique against five baselines from two categories: traditional IR-based techniques~\cite{saha_bleuir, blizzard} and LLM-based approaches~\cite{BRaIn, Agentless, LocAgent}. For comparison, we use the CogniGent (Devstral) variant, which offers a balanced trade-off between performance, computational cost, and execution time (Table \ref{tab:performance_RQ1}).

We compare our technique with traditional IR-based techniques: Baseline IR \cite{ir_localization_lda_buglocator} and two techniques from the literature, BLIZZARD~\cite{blizzard} and BLUiR~\cite{saha_bleuir}. To replicate the Baseline IR, we extract all methods and constructors from the source repository and index them using Apache Lucene~\cite{lucene}, while keeping the default BM25 parameters ($k$ and $b$)~\cite{robertson1995okapi_bm25} and the preprocessing settings. During localization, we use bug reports (title + description) as queries to retrieve buggy code segments from the target system and version, as reported. To replicate BLIZZARD~\cite{blizzard} and BLUiR~\cite{saha_bleuir}, we follow a similar indexing policy and index the source code in Apache Lucene~\cite{lucene}. BLIZZARD classifies bug reports into three types and uses PageRank-based \cite{pagerank} text graph to generate tailored queries to retrieve the buggy code segments. On the other hand, BLUiR constructs structured queries from bug reports and source elements, performing eight separate searches over class names, method names, variable names, comments, and bug report fields (title and description). The results of these searches are then combined to produce an overall suspiciousness score for localizing buggy segments.  We collected the BLUiR replication package from Bench4BL \cite{bench4bl} and the BLIZZARD implementation from its replication package~\cite{BLIZZARD_github, bench4bl}, and adapted both for version-based retrieval.

\begin{table*}[!ht]
    \caption{Comparison of Baselines against CogniGent\vspace{-.4cm}}
    \centering
    \small
    \begin{tabular}{|c|
        c|c||  
        c|c||  
        c|c||| 
        c|c||  
        c|c||  
        c|c||  
        c|c|}  
        \hline
        \multirow{3}{*}{\textbf{Metrics}} 
            & \multicolumn{6}{c|||}{\textbf{IR-based Techniques}} 
            & \multicolumn{8}{c|}{\textbf{LLM-based Techniques}} \\ \cline{2-15}
            & \multicolumn{2}{c||}{Baseline-IR} 
            & \multicolumn{2}{c||}{BLUiR} 
            & \multicolumn{2}{c|||}{BLIZZARD}
            & \multicolumn{2}{c||}{BRaIn} 
            & \multicolumn{2}{c||}{Agentless} 
            & \multicolumn{2}{c||}{LocAgent} 
            & \multicolumn{2}{c|}{CogniGent} \\ \cline{2-15}
            & Doc. & Met. 
            & Doc. & Met. 
            & Doc. & Met.
            & Doc. & Met. 
            & Doc. & Met. 
            & Doc. & Met.
            & Doc. & Met. \\ \hline\hline

        MAP   & 0.330 & 0.163 & 0.344 & 0.191 & 0.380 & 0.177 
               & 0.375 & 0.206 & 0.332 & 0.177 & 0.384 & 0.207 & 0.407 & 0.226 \\ \hline
        MRR   & 0.334 & 0.165 & 0.350 & 0.196 & 0.387 & 0.183 
               & 0.395 & 0.223 & 0.352 & 0.176 & 0.392 & 0.224 & 0.418 & 0.254 \\ \hline
        HIT@1 & 0.254 & 0.122 & 0.257 & 0.139 & 0.321 & 0.127 
               & 0.343 & 0.187 & 0.324 & 0.137 & 0.353 & 0.197 & 0.374 & 0.227 \\ \hline
        HIT@5 & 0.443 & 0.217 & 0.453 & 0.259 & 0.446 & 0.234 
               & 0.464 & 0.267 & 0.448 & 0.217 & 0.465 & 0.265 & 0.475 & 0.272 \\ \hline
    \end{tabular}
    \vspace{2mm}
    
    {\footnotesize \textit{Doc.: Document-level localization, Met.: Method-level localization.}}
    \label{tab:RQ3_baseline_comparison}
    \vspace{-0.3cm}
\end{table*}

We also compare CogniGent against three LLM-based techniques- BRaIn~\cite{BRaIn}, Agentless~\cite{Agentless} and LocAgent~\cite{LocAgent}. BRaIn is a state of the art technique that improves bug localization through Intelligent Relevance Feedback. It captures contextual feedback from LLMs on initially retrieved documents, leverages them to enhance the query, and then employs Information Retrieval to localize bugs. For the experiment, we collected the authors’ replication package \cite{replication_BRaIn}. AgentLess is a recent technique that combines LLM-based reasoning with embedding-based retrieval to identify buggy source documents, which are then used for fixing corresponding issues. It builds a structured repository representation, retrieves semantically related documents using embeddings, and then uses the LLM to reason over their skeleton forms functions to refine and rank the most likely buggy locations. We adapt the replication package from the authors~\cite{agentless_replication} and use the localization portion of the technique for the experiment. On the other hand, LocAgent is a graph-guided agentic framework for issue localization. It builds a heterogeneous code graph of a project and creates text indexes (e.g., BM25) over code entities to quickly retrieve those relevant to a bug report. The LLM agent then searches and traverses the graph, reasoning over the retrieved entities to produce a ranked list of suspicious documents and functions. We collected the replication packages from the authors~\cite{locagent_replication} and adapted them for our dataset and version-based localization and comparison.

\begin{figure}[t]  
    \centering
    \includegraphics[width=0.8\linewidth]{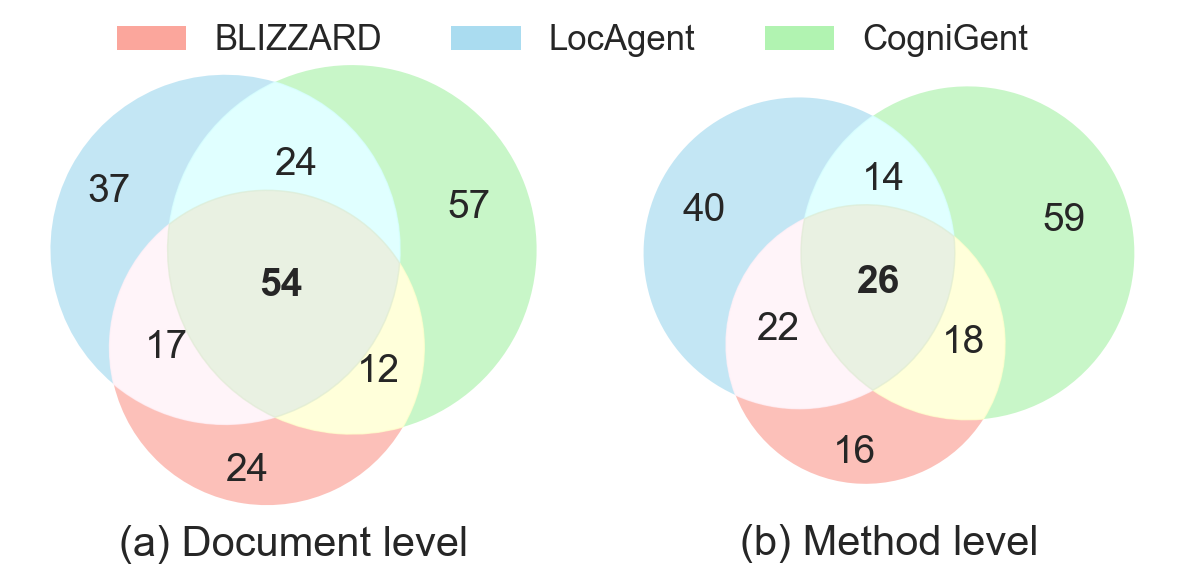}  
    \vspace{-0.2cm}
    \caption{\centering Coverage of Buggy Components Localized within Top-5 Across All Multi-Component Bugs}
    \label{fig:RQ3b_multi-bug_performance}
    \vspace{-0.5cm}
\end{figure}

Table~\ref{tab:RQ3_baseline_comparison} summarizes the comparison results between CogniGent and six baseline methods. Among traditional techniques, BLUiR achieves a MAP of 0.344, while BLIZZARD and the baseline-IR reach 0.379 and 0.330, respectively. CogniGent outperforms all of them with a MAP of 0.407, marking up to 23.33\% improvement, and achieves a maximum 25.14\% gain in MRR (0.418). It also improves HIT@1 and HIT@5 to 0.374 and 0.475, yielding 16.31–47.33\% and 5.00–7.25\% gains against IR baselines in these metrics, respectively.
At the method level, CogniGent achieves a MAP of 0.226 compared to 0.163, 0.191, and 0.177 for Baseline-IR, BLUiR, and BLIZZARD, showing 18.32–38.57\% improvements. In MRR, HIT@1, and HIT@5, it also performs well, with gains of up to 53.74\%, 86.11\%, and 25.78\%. These results demonstrate our technique's benefits over the traditional baseline techniques.

CogniGent also performs better than LLM–based techniques (Table~\ref{tab:RQ3_baseline_comparison}), which capture relationships between bug reports and faulty code through reasoning. At the document level, CogniGent achieves a MAP score of 0.407, representing improvements of 8.53\%, 22.59\%, and 5.98\% over BRaIn, AgentLess, and LocAgent, respectively. Similar gains are observed in MRR, HIT@1, and HIT@5, with improvements of up to 18.75\%, 15.46\%, and 6.20\%.
At the method level, CogniGent further surpasses these techniques, achieving MAP and MRR scores of 0.226 and 0.254—maximum gains of 31.39\% and 43.50\%, respectively. For HIT@1 and HIT@5, it records improvements of 65.60\% and 25.58\%. Overall, these results demonstrate CogniGent’s effectiveness in ranking buggy documents closer at higher positions.

We further examine CogniGent’s performance on multi-component bugs, where a single bug affects multiple documents or methods. Fig.~\ref{fig:RQ3b_multi-bug_performance} compares CogniGent with BLIZZARD and LocAgent, the closest competitors. In our dataset, we have 185 bug reports, each of which triggers changes across multiple documents, totaling 523 buggy documents. Among these CogniGent localized 147 (28.10\%) in the top-5 positions, outperforming LocAgent (132) and BLIZZARD (107). All three techniques commonly localized 54 documents, while CogniGent uniquely localized 57, compared to 37 by LocAgent and 24 by BLIZZARD. At the method level, 328 bug reports target 1,766 buggy methods in our dataset, where CogniGent localized 117 methods in the top-5, again outperforming LocAgent (102) and BLIZZARD (82). These results confirm that even at finer granularity, CogniGent provides consistently superior coverage of multi-component bugs.

To further validate these performance differences, we conducted non-parametric statistical tests - Wilcoxon signed-rank test \cite{wilcoxon_statistical} and measured effect size with Cliff's $\delta$ \cite{wilcoxon_statistical} using the ranks of first correct results from each three techniques above. As shown in Table~\ref{tab:statistical_test_comparison}, CogniGent demonstrates statistically significant improvements over BLIZZARD at both Top-1 and Top-5 positions ($p < 0.05$), with medium to large effect sizes. When compared to LocAgent, the results also show statistically significant improvements, with small to medium effect sizes. These findings confirm that CogniGent consistently localizes buggy components more effectively than both BLIZZARD and LocAgent.

 \vspace{-0.1cm}
 \begin{tcolorbox}[colback=lightergray,colframe=black,arc=1mm,boxrule=0pt,
     leftrule=2pt, bottomrule=1pt]
 \textbf{RQ3 Summary:} CogniGent outperforms both traditional and LLM-based baselines, achieving 23.33\% and 38.57\% higher MAP scores at the document and method levels, respectively. This suggests that dynamic cognitive debugging may have contributed to CogniGent localizing buggy documents closer to the top, with observed improvements of 2.36–86.11\% across metrics compared to techniques that do not consider inter-component relationships. Statistical significance tests further confirm its superiority.
 \end{tcolorbox}

 \vspace{-0.2cm}

\begin{table}[t]
	\caption{Statistical Test: CogniGent vs. BLIZZARD and LocAgent}
	\vspace{-0.2cm}
	\centering
	\begin{tabular}{|l|c|c|}
		\hline
		Evaluation Point & \textit{vs.} BLIZZARD & \textit{vs}. LocAgent \\ 
		\hline \hline
		Top-1  & 0.0027 **, (Medium$^\dagger$) & 0.0180 *, (Small$^\dagger$) \\ \hline
		Top-5  & 0.0011 **, (Large$^\dagger$)  & 0.0082 **, (Medium$^\dagger$) \\ \hline 
	\end{tabular}
	\label{tab:statistical_test_comparison}
	\vspace{-1pt} 
	{\footnotesize
		$*$ = statistical significance, \quad $^\dagger$ =  Effect size (Cliff’s $\delta$)
	}
	\vspace{-0.5cm}
\end{table}

\section{Related Work}
\textbf{IR and Bug Localization.} Traditional bug localization methods fall into two categories: spectra-based and Information Retrieval (IR)-based~\cite{ir_bug_localization1}. Spectra-based approaches rely on execution traces and test cases, making them resource-intensive and often impractical~\cite{spectra2}. On the other hand, IR-based methods treat bug localization as a retrieval task and rely on the textual similarity between bug reports and source code to identify buggy documents.

Early IR-based techniques used the Vector Space Model (VSM)~\cite{vector_space_model}. Later work refined these approaches with contextual cues such as bug resolution history, code changes, and version metadata~\cite{saha2014effectiveness_bug_rep_history, wen2016locus, sisman2012incorporating}. Saha et al.~\cite{saha_bleuir} enhanced retrieval using structured features via the Indri engine~\cite{Indri}, while BugLocator~\cite{ir_localization_lda_buglocator} combined rVSM scores with bug-fix history. AmaLgam and AmaLgam+~\cite{ir_ml_amalgam, wang_amalgam+} integrated multiple IR techniques (e.g., BLUiR, BugLocator) with stack traces and version or reporter history. Advanced models like LSI and LDA~\cite{ir_localization_lda_buglocator, ir_bug_localization4} capture latent topics to address vocabulary mismatch; however, their performance remains comparable to simpler VSM-based approaches~\cite{bench4bl}.

Since IR performance largely depends on query quality, some researchers focused on query reformulation—refining bug report queries by adding or removing terms~\cite{mills2020relationship, rahman2021forgotten}. Refoqus~\cite{prf_haiduc_4} employed machine learning to suggest expansion or reduction strategies, while graph-based and genetic methods~\cite{blizzard, rahman2021forgotten} optimized queries by analyzing term relationships. Relevance feedback techniques, such as Rocchio’s algorithm~\cite{prf_haiduc_4} and Spatial Code Proximity (SCP)~\cite{prf_sisman_kak_1}, further reformulate queries based on statistical and co-occurrence patterns from initial search results.

These traditional approaches rely on textual matching and statistical cues. While we also apply BM25-based retrieval (e.g., Lucene in Neo4j) to gather initial candidates, we perform multi-step LLM reasoning to capture symptom-level cues and align them with code behavior, moving beyond surface-level similarity.

\textbf{Deep Learning and Bug Localization.} Advances in deep learning and Large Language Models (LLMs) have improved the understanding of source code and natural language, motivating their application to bug localization. Existing approaches can be broadly categorized into training-based models and prompt-based instruction-following methods.

Among training-based techniques, DNNLOC~\cite{dnnloc_ir}, a seminal work, trains on positive and negative (bug report, source file) pairs and incorporates multiple signals (e.g., rVSM similarity~\cite{ir_localization_lda_buglocator}, class-name similarity, bug-fix recency), with a second network combining them for ranking. However, its dependence on bug-fixing recency can limit practical use. FBL-BERT~\cite{ciborowska_fbl_bert} trains on bug-report and changeset pairs and follows ColBERT~\cite{khattab2020colbert}-style late interaction for ranking, matching bug-report tokens to their most similar tokens in each changeset to identify buggy code. However, its reliance on changesets limits its applicability in rapidly evolving projects. BLAZE~\cite{blaze} trains a GPT-based model using hard-example contrastive learning and dynamic chunking to improve alignment between bug reports and source code, enabling effective retrieval of buggy documents. Other approaches have employed convolutional neural networks to learn representations for bug localization~\cite{TRANP-CNN, cdnn_2023}. However, these techniques can lack generalizability and require retraining, limiting scalability in large, evolving codebases.

Prompt-based approaches leverage LLM reasoning to detect software bugs. BRaIn applies Intelligent Relevance Feedback, determining document relevance leveraging an LLM, making better search queries, and refining document retrieval through re-ranking. LocAgent~\cite{LocAgent} uses an agentic workflow where an LLM performs reasoning over a heterogeneous code graph. Guided by BM25-based retrieval and graph traversal, the agent iteratively hops across nodes (e.g., files, functions) through multiple edge types to identify likely buggy components. CoSIL~\cite{CoSIL} similarly performs graph-guided localization through a two-stage process: expanding the search space via module-level call graphs and refining to function-level candidates via iterative pruning.

In contrast, our technique leverages the pre-trained knowledge of LLMs while retrieving only a limited set of code segments via Lucene~\cite{lucene}, improving scalability. Through code navigation guided by human-inspired causal reasoning and maintaining a focused context window, CogniGent gains deeper understanding of root causes and improves bug localization.

\section{Threats to Validity}
Threats to internal validity, which involve potential experimental errors or biases, primarily arise from replicating existing baselines. To address this, we relied on replication packages released by the original authors (e.g., BLIZZARD, BRaIn, Agentless, LocAgent~\cite{BLIZZARD_github, Agentless, BRaIn, LocAgent}) and from Bench4BL~\cite{bench4bl} (BLUiR~\cite{saha_bleuir}). Since Indri~\cite{Indri} is now deprecated, we replaced it with Lucene in our BLUiR replication. To reduce bias, we tested on two datasets and found minimal differences from the baselines.

Threats to external validity pertain to the generalizability of our findings. Although CogniGent was evaluated solely on Java projects, the underlying language models (e.g., LLaMA~\cite{llama3herd2024}) are inherently adaptable to multiple programming languages, which helps mitigate this limitation.

Threats to construct validity concern the suitability of the evaluation metrics. We adopted widely accepted measures—Mean Average Precision (MAP), Mean Reciprocal Rank (MRR), and HIT@K—which are standard in both bug localization~\cite{blizzard, saha_bleuir, Rack} and Information Retrieval research~\cite{evaluation_ir}. Using these established metrics helps ensure the reliability of our evaluation.

Finally, we opted for locally deployable, open-weight models (e.g., LLaMA~\cite{llama3herd2024}) for our experiments, as reproducibility with proprietary models (e.g., GPT-4~\cite{openai_gpt4}) might be affected by cost barriers. Additionally, we release the replication package~\cite{cognigent_replication}, which mitigates the risks associated with experimental configurations (e.g., temperature = 0.5), but results may vary due to LLM non-determinism~\cite{stochasticity_LLM}.

\section{Conclusion}

Software bugs undermine system reliability and security, leading to significant financial losses~\cite{atAndT_1, BI_crowdstrike}. To address these long-standing challenges, we introduce \textit{CogniGent}, a technique designed to overcome the limitations of IR and AI-based bug localization by enabling causal reasoning, dependency-aware navigation, and context-managed code exploration. CogniGent applies human-inspired Dynamic Cognitive Debugging within an LLM-driven agentic workflow to formulate hypotheses, traverse interconnected components, and trace fault propagation to identify underlying causes. To evaluate CogniGent’s performance, we use three widely adopted metrics in bug localization--Mean Average Precision (MAP), Mean Reciprocal Rank (MRR), and HIT@K—and observed substantial improvements: 23.33\% and 38.57\% gains in MAP at the document and method levels, respectively, and 25.14\% and 53.74\% gains in MRR across the same levels of granularity. Building on these results, we plan to capture finer-grained structural relationships and extend our agentic workflow to further support bug resolution.

\vspace{-0.2cm}
\begin{acks}
    This work was supported by the Natural Sciences and Engineering Research Council of Canada (Discovery Grant RGPIN‑03236)
\end{acks}
\vspace{-0.2cm}


\bibliographystyle{ACM-Reference-Format}
\bibliography{bibliography}

@misc{atAndT_1,
  author    = {Zahn, Morgan and Haworth, Jon and Margolin, Josh and Date, Jack and Barr, Luke},
  title     = {{AT\&T} outage caused by software update, company says},
  year      = {2024},
  month     = {February},
  day       = {22},
  url       = {https://abcnews.go.com/US/att-outage-impacting-us-customers-company/story?id=107440297},
  note      = {Accessed: 2025-09-22},
  publisher = {ABC News}
}

@misc{atAndT_2,
  author       = {{Federal Communications Commission}},
  title        = {February 22, 2024 {{AT\&T}} Mobility Network Outage Report and Findings},
  year         = {2024},
  month        = {July},
  url          = {https://www.benton.org/headlines/february-22-2024-att-mobility-network-outage-report-and-findings},
  note         = {Published by the Benton Institute for Broadband \& Society. Details the cause and impact of a nationwide {{AT\&T}} wireless service outage affecting over 125 million devices and blocking more than 92 million calls.}
}

@misc{BI_crowdstrike, title={Here comes the wave of insurance claims for the CrowdStrike outage}, url={https://shorturl.at/5Y1jQ}, journal={Business Insider}, author={Wee, Lian Kit}, year={2024}, month=jul, language={en-US} }

@article{o2017debugging_stats,
  title={The Debugging Mindset: Understanding the psychology of learning strategies leads to effective problem-solving skills.},
  author={O'Dell, Devon H},
  journal={Queue},
  volume={15},
  number={1},
  pages={71--90},
  year={2017},
  publisher={ACM New York, NY, USA}
}

@article{britton2013reversible_stats,
  title={Reversible debugging software},
  author={Britton, Tom and Jeng, Lisa and Carver, Graham and Cheak, Paul and Katzenellenbogen, Tomer},
  journal={Judge Bus. School, Univ. Cambridge, Cambridge, UK, Tech. Rep},
  volume={229},
  year={2013}
}

@misc{devops2024_dev_stats,
  author    = {DevOps},
  title     = {Survey: Fixing Bugs Stealing Time from Development},
  year      = {2024},
  url       = {https://shorturl.at/Fj8sB},
  
}

@inproceedings{blizzard,
  title={Improving ir-based bug localization with context-aware query reformulation},
  author={Rahman, Mohammad Masudur and Roy, Chanchal K},
  booktitle={Proceedings of the 2018 26th ACM joint meeting on European software engineering conference and symposium on the foundations of software engineering},
  pages={621--632},
  year={2018}
}

@misc{BLIZZARD_github,
  author       = {Masud Rahman},
  title        = {BLIZZARD},
  url          = {https://github.com/masud-technope/BLIZZARD}
}

@article{rahman2021forgotten,
  title={The forgotten role of search queries in ir-based bug localization: an empirical study},
  author={Rahman, Mohammad Masudur and Khomh, Foutse and Yeasmin, Shamima and Roy, Chanchal K},
  journal={Empirical Software Engineering},
  volume={26},
  number={6},
  pages={116},
  year={2021},
  publisher={Springer}
}

@article{furnas1987vocabulary,
  title={The vocabulary problem in human-system communication},
  author={Furnas, George W. and Landauer, Thomas K. and Gomez, Louis M. and Dumais, Susan T.},
  journal={Communications of the ACM},
  volume={30},
  number={11},
  pages={964--971},
  year={1987},
  publisher={ACM New York, NY, USA}
}

@inproceedings{ir_bug_localization1,
  title={Evaluating the usefulness of IR-based fault localization techniques},
  author={Wang, Qianqian and Parnin, Chris and Orso, Alessandro},
  booktitle={Proceedings of the 2015 International Symposium on Software Testing and Analysis},
  pages={1--11},
  year={2015}
}

@inproceedings{ir_bug_localization2_spectra,
  title={Information retrieval and spectrum based bug localization: Better together},
  author={Le, Tien-Duy B and Oentaryo, Richard J and Lo, David},
  booktitle={Proceedings of the 2015 10th Joint Meeting on Foundations of Software Engineering},
  pages={579--590},
  year={2015}
}

@inproceedings{ir_bug_localization3_topic,
  title={A topic-based approach for narrowing the search space of buggy files from a bug report},
  author={Nguyen, Anh Tuan and Nguyen, Tung Thanh and Al-Kofahi, Jafar and Nguyen, Hung Viet and Nguyen, Tien N},
  booktitle={2011 26th IEEE/ACM International Conference on Automated Software Engineering (ASE 2011)},
  pages={263--272},
  year={2011},
  organization={IEEE}
}

@article{ir_bug_localization4,
  title={Feature location using probabilistic ranking of methods based on execution scenarios and information retrieval},
  author={Poshyvanyk, Denys and Guneuc, Yann-Gal and Marcus, Andrian and Antoniol, Giuliano and Rajlich, Vaclav},
  journal={IEEE Transactions on Software Engineering},
  volume={33},
  number={6},
  pages={420--432},
  year={2007},
  publisher={IEEE}
}

@inproceedings{ir_ml_amalgam,
  title={Version history, similar report, and structure: Putting them together for improved bug localization},
  author={Wang, Shaowei and Lo, David},
  booktitle={Proceedings of the 22nd International Conference on Program Comprehension},
  pages={53--63},
  year={2014}
}

@article{wang_amalgam+,
  title={Amalgam+: Composing rich information sources for accurate bug localization},
  author={Wang, Shaowei and Lo, David},
  journal={Journal of Software: Evolution and Process},
  volume={28},
  number={10},
  pages={921--942},
  year={2016},
  publisher={Wiley Online Library}
}

@article{query_reformulation_tosem,
  title={A Systematic Review of Automated Query Reformulations in Source Code Search},
  author={Rahman, Mohammad Masudur and Roy, Chanchal K},
  journal={ACM Transactions on Software Engineering and Methodology},
  year={2021},
  publisher={ACM New York, NY}
}

@article{ir_localization_lda_buglocator,
  title={Where should we fix this bug? a two-phase recommendation model},
  author={Kim, Dongsun and Tao, Yida and Kim, Sunghun and Zeller, Andreas},
  journal={IEEE transactions on software Engineering},
  volume={39},
  number={11},
  pages={1597--1610},
  year={2013},
  publisher={IEEE}
}

@inproceedings{ir_localization_ml_dl,
  title={Bug localization with combination of deep learning and information retrieval},
  author={Lam, An Ngoc and Nguyen, Anh Tuan and Nguyen, Hoan Anh and Nguyen, Tien N},
  booktitle={2017 IEEE/ACM 25th International Conference on Program Comprehension (ICPC)},
  pages={218--229},
  year={2017},
  organization={IEEE}
}

@inproceedings{chaparro2017_ob_eb,
  title={Using observed behavior to reformulate queries during text retrieval-based bug localization},
  author={Chaparro, Oscar and Florez, Juan Manuel and Marcus, Andrian},
  booktitle={2017 IEEE International Conference on Software Maintenance and Evolution (ICSME)},
  pages={376--387},
  year={2017},
  organization={IEEE}
}

@inproceedings{sisman2012incorporating,
  title={Incorporating version histories in information retrieval based bug localization},
  author={Sisman, Bunyamin and Kak, Avinash C},
  booktitle={2012 9th IEEE working conference on mining software repositories (MSR)},
  pages={50--59},
  year={2012},
  organization={IEEE}
}

@article{lucene,
  title = {Apache Lucene},
  author = {The Apache Software Foundation},
  year = {2021},
  url = {https://lucene.apache.org/},
}

@article{practitioners_bug_localization_study,
  title={How practitioners perceive automated bug report management techniques},
  author={Zou, Weiqin and Lo, David and Chen, Zhenyu and Xia, Xin and Feng, Yang and Xu, Baowen},
  journal={IEEE Transactions on Software Engineering},
  volume={46},
  number={8},
  pages={836--862},
  year={2018},
  publisher={IEEE}
}

@inproceedings{bench4bl,
  title={Bench4bl: reproducibility study on the performance of ir-based bug localization},
  author={Lee, Jaekwon and Kim, Dongsun and Bissyand{\'e}, Tegawend{\'e} F and Jung, Woosung and Le Traon, Yves},
  booktitle={Proceedings of the 27th ACM SIGSOFT international symposium on software testing and analysis},
  pages={61--72},
  year={2018}
}

@article{robertson1995okapi_bm25,
  title={Okapi at TREC-3},
  author={Robertson, Stephen E and Walker, Steve and Jones, Susan and Hancock-Beaulieu, Micheline M and Gatford, Mike and others},
  journal={Nist Special Publication Sp},
  volume={109},
  pages={109},
  year={1995},
  publisher={National Instiute of Standards \& Technology}
}

@inproceedings{saha_bleuir,
  title={Improving bug localization using structured information retrieval},
  author={Saha, Ripon K and Lease, Matthew and Khurshid, Sarfraz and Perry, Dewayne E},
  booktitle={2013 28th IEEE/ACM International Conference on Automated Software Engineering (ASE)},
  pages={345--355},
  year={2013},
  organization={IEEE}
}

@article{mills2020relationship,
  title={On the relationship between bug reports and queries for text retrieval-based bug localization},
  author={Mills, Chris and Parra, Esteban and Pantiuchina, Jevgenija and Bavota, Gabriele and Haiduc, Sonia},
  journal={Empirical Software Engineering},
  volume={25},
  pages={3086--3127},
  year={2020},
  publisher={Springer}
}

@ARTICLE{dreamloc,
  author={Qi, Binhang and Sun, Hailong and Yuan, Wei and Zhang, Hongyu and Meng, Xiangxin},
  journal={IEEE Transactions on Reliability}, 
  title={DreamLoc: A Deep Relevance Matching-Based Framework for bug Localization}, 
  year={2022},
  volume={71},
  number={1},
  pages={235-249},
  doi={10.1109/TR.2021.3104728}}

@INPROCEEDINGS{dnnloc_ir,
  author={Lam, An Ngoc and Nguyen, Anh Tuan and Nguyen, Hoan Anh and Nguyen, Tien N.},
  booktitle={2017 IEEE/ACM 25th International Conference on Program Comprehension (ICPC)}, 
  title={Bug Localization with Combination of Deep Learning and Information Retrieval}, 
  year={2017},
  volume={},
  number={},
  pages={218-229},
  doi={10.1109/ICPC.2017.24}}

@inproceedings{ciborowska_fbl_bert,
  title={Fast changeset-based bug localization with BERT},
  author={Ciborowska, Agnieszka and Damevski, Kostadin},
  booktitle={Proceedings of the 44th International Conference on Software Engineering},
  pages={946--957},
  year={2022}
}

@article{pagerank,
  title={The anatomy of a large-scale hypertextual web search engine},
  author={Brin, Sergey and Page, Lawrence},
  journal={Computer networks and ISDN systems},
  volume={30},
  number={1-7},
  pages={107--117},
  year={1998},
  publisher={Elsevier}
}

@INPROCEEDINGS{bl_code_change,
  author={Youm, Klaus Changsun and Ahn, June and Kim, Jeongho and Lee, Eunseok},
  booktitle={2015 Asia-Pacific Software Engineering Conference (APSEC)}, 
  title={Bug Localization Based on Code Change Histories and Bug Reports}, 
  year={2015},
  volume={},
  number={},
  pages={190-197},
  doi={10.1109/APSEC.2015.23}}

@article{summarization,
  title={A transformer-based approach for source code summarization},
  author={Ahmad, Wasi Uddin and Chakraborty, Saikat and Ray, Baishakhi and Chang, Kai-Wei},
  journal={arXiv preprint arXiv:2005.00653},
  year={2020}
}

@ARTICLE{vector_space_model,
  author={Lee, D.L. and Huei Chuang and Seamons, K.},
  journal={IEEE Software}, 
  title={Document ranking and the vector-space model}, 
  year={1997},
  volume={14},
  number={2},
  pages={67-75},
  doi={10.1109/52.582976}}

@inproceedings{Indri,
  title={Indri : A language-model based search engine for complex queries ( extended version )},
  author={Trevor Strohman and Donald Metzler and Howard R. Turtle and W. Bruce Croft},
  year={2005},
  url={https://api.semanticscholar.org/CorpusID:18471028}
}

@inproceedings{code_generation,
  title={Intellicode compose: Code generation using transformer},
  author={Svyatkovskiy, Alexey and Deng, Shao Kun and Fu, Shengyu and Sundaresan, Neel},
  booktitle={Proceedings of the 28th ACM joint meeting on European software engineering conference and symposium on the foundations of software engineering},
  pages={1433--1443},
  year={2020}
}

@inproceedings{spectra2,
author = {Wang, Qianqian and Parnin, Chris and Orso, Alessandro},
title = {Evaluating the usefulness of IR-based fault localization techniques},
year = {2015},
isbn = {9781450336208},
publisher = {Association for Computing Machinery},
address = {New York, NY, USA},
url = {https://doi.org/10.1145/2771783.2771797},
doi = {10.1145/2771783.2771797},
booktitle = {Proceedings of the 2015 International Symposium on Software Testing and Analysis},
pages = {1–11},
numpages = {11},
location = {Baltimore, MD, USA},
series = {ISSTA 2015}
}

@inproceedings{saha2014effectiveness_bug_rep_history,
  title={On the effectiveness of information retrieval based bug localization for c programs},
  author={Saha, Ripon K and Lawall, Julia and Khurshid, Sarfraz and Perry, Dewayne E},
  booktitle={2014 IEEE international conference on software maintenance and evolution},
  pages={161--170},
  year={2014},
  organization={IEEE}
}

@inproceedings{wen2016locus,
  title={Locus: Locating bugs from software changes},
  author={Wen, Ming and Wu, Rongxin and Cheung, Shing-Chi},
  booktitle={Proceedings of the 31st IEEE/ACM International Conference on Automated Software Engineering},
  pages={262--273},
  year={2016}
}

@book{evaluation_ir,
  title={Information retrieval evaluation},
  author={Harman, Donna},
  year={2011},
  publisher={Morgan \& Claypool Publishers}
}

@inproceedings{Rack,
   author = {Mohammad Masudur Rahman and Chanchal K. Roy and David Lo},
   doi = {10.1109/saner.2016.80},
   month = {5},
   pages = {349-359},
   publisher = {Institute of Electrical and Electronics Engineers (IEEE)},
   title = {RACK: Automatic API Recommendation Using Crowdsourced Knowledge},
   year = {2016},
}

@INPROCEEDINGS{prf_sisman_kak_1,
  author={Sisman, Bunyamin and Kak, Avinash C.},
  booktitle={2013 10th Working Conference on Mining Software Repositories (MSR)}, 
  title={Assisting code search with automatic Query Reformulation for bug localization}, 
  year={2013},
  volume={},
  number={},
  pages={309-318},
  doi={10.1109/MSR.2013.6624044}
}

@inproceedings{prf_haiduc_4,
  title={Automatic query reformulations for text retrieval in software engineering},
  author={Haiduc, Sonia and Bavota, Gabriele and Marcus, Andrian and Oliveto, Rocco and De Lucia, Andrea and Menzies, Tim},
  booktitle={2013 35th International Conference on Software Engineering (ICSE)},
  pages={842--851},
  year={2013},
  organization={IEEE}
}

@article{Agentless,
  title={Agentless: Demystifying llm-based software engineering agents},
  author={Xia, Chunqiu Steven and Deng, Yinlin and Dunn, Soren and Zhang, Lingming},
  journal={arXiv preprint arXiv:2407.01489},
  year={2024}
}

@inproceedings{TRANP-CNN,
  title={Enhancing the Unified Features to Locate Buggy Files by Exploiting the Sequential Nature of Source Code.},
  author={Huo, Xuan and Li, Ming},
  booktitle={IJCAI},
  pages={1909--1915},
  year={2017}
}

@inproceedings{khattab2020colbert,
  title={Colbert: Efficient and effective passage search via contextualized late interaction over bert},
  author={Khattab, Omar and Zaharia, Matei},
  booktitle={Proceedings of the 43rd International ACM SIGIR conference on research and development in Information Retrieval},
  pages={39--48},
  year={2020}
}

@misc{javaparser,
  title={Javaparser},
  author={Fischer, Christoph},
  year={2019},
  url={https://github.com/javaparser/javaparser}
}

@book{wilcoxon_statistical,
  title={Testing Significance Tests: A Simulation with Cliff's Delta, t-tests, and Mann-Whitney U},
  author={Barnes, Tyler and Moore, Scott C. and Osatuke, Katerine},
  year={2018},
  publisher={National Center for Organizational Development, Department of Veteran Affairs},
  isbn={978-0-692-86421-3}
}

@misc{replication_BRaIn,
  author = {Asif Samir, Mohammad Masudur Rahman},
  title = {{BRaIn: Replication Package}},
  year = {2024},
  howpublished = {\url{https://github.com/asifsamir/BRaIn}}
}

@techreport{tricentis2025quality,
  title     = {2025 Quality Transformation Report},
  author    = {{Tricentis}},
  year      = {2025},
  institution = {Tricentis},
  url       = {https://www.tricentis.com/resources/quality-transformation-report},
  note      = {Accessed: 2025-09-25}
}

@article{debugging_study,
  title={What constitutes debugging? An exploratory study of debugging episodes},
  author={Alaboudi, Abdulaziz and LaToza, Thomas D},
  journal={Empirical Software Engineering},
  volume={28},
  number={5},
  pages={117},
  year={2023},
  publisher={Springer}
}

@article{BRaIn,
  title={Improved IR-based Bug Localization with Intelligent Relevance Feedback},
  author={Samir, Asif Mohammed and Rahman, Mohammad Masudur},
  journal={arXiv preprint arXiv:2501.10542},
  year={2025}
}

@article{LocAgent,
  title={Locagent: Graph-guided llm agents for code localization},
  author={Chen, Zhaoling and Tang, Xiangru and Deng, Gangda and Wu, Fang and Wu, Jialong and Jiang, Zhiwei and Prasanna, Viktor and Cohan, Arman and Wang, Xingyao},
  journal={arXiv preprint arXiv:2503.09089},
  year={2025}
}

@article{codeXgraph,
   author = {Xiangyan Liu and Bo Lan and Zhiyuan Hu and Yang Liu and Zhicheng Zhang and Fei Wang and Michael Shieh and Wenmeng Zhou},
   month = {8},
   title = {CodexGraph: Bridging Large Language Models and Code Repositories via Code Graph Databases},
   url = {http://arxiv.org/abs/2408.03910},
   year = {2024},
}

@article{sweBench,
  title={Swe-bench: Can language models resolve real-world github issues?},
  author={Jimenez, Carlos E and Yang, John and Wettig, Alexander and Yao, Shunyu and Pei, Kexin and Press, Ofir and Narasimhan, Karthik},
  journal={arXiv preprint arXiv:2310.06770},
  year={2023}
}

@inproceedings{AutoCodeRover,
  title={Autocoderover: Autonomous program improvement},
  author={Zhang, Yuntong and Ruan, Haifeng and Fan, Zhiyu and Roychoudhury, Abhik},
  booktitle={Proceedings of the 33rd ACM SIGSOFT International Symposium on Software Testing and Analysis},
  pages={1592--1604},
  year={2024}
}

@article{CoSIL,
  title={CoSIL: Software Issue Localization via LLM-Driven Code Repository Graph Searching},
  author={Jiang, Zhonghao and Ren, Xiaoxue and Yan, Meng and Jiang, Wei and Li, Yong and Liu, Zhongxin},
  journal={arXiv preprint arXiv:2503.22424},
  year={2025}
}

@misc{codeNL_semanticSearch_fail_1,
  author    = {{Greptile}},
  title     = {Semantic Codebase Search: Why It’s Hard and What Works},
  year      = {2023},
  url       = {https://www.greptile.com/blog/semantic-codebase-search},
  note      = {Accessed: 2025-09-25},
  publisher = {Greptile}
}

@misc{codeNL_semanticSearch_fail_2,
  author    = {{GitHub}},
  title     = {Towards Natural Language Semantic Code Search},
  year      = {2024},
  url       = {https://github.blog/ai-and-ml/machine-learning/towards-natural-language-semantic-code-search/},
  note      = {Accessed: 2025-09-25},
  publisher = {GitHub}
}

@article{context_confusion_1,
  title={Llms can be easily confused by instructional distractions},
  author={Hwang, Yerin and Kim, Yongil and Koo, Jahyun and Kang, Taegwan and Bae, Hyunkyung and Jung, Kyomin},
  journal={arXiv preprint arXiv:2502.04362},
  year={2025}
}

@book{cognitive_debugging_book,
  author    = {Leite, J. C. S. do Prado and Travassos, G. H.},
  title     = {Cognitive Debugging},
  year      = {2023},
  publisher = {Springer},
  isbn      = {9783031420637},
  url       = {https://link.springer.com/book/10.1007/978-3-031-42064-1},
  note      = {Accessed: 2025-09-25}
}

@techreport{context_rot_context_length_impact,
  title        = {Context Rot: How Increasing Input Tokens Impacts LLM Performance},
  author       = {Hong, Kelly and Troynikov, Anton and Huber, Jeff},
  year         = {2025},
  month        = {July},
  institution  = {Chroma},
  url          = {https://research.trychroma.com/context-rot}
}

@misc{context_length_impact_1,
      title={Same Task, More Tokens: the Impact of Input Length on the Reasoning Performance of Large Language Models}, 
      author={Mosh Levy and Alon Jacoby and Yoav Goldberg},
      year={2024},
      eprint={2402.14848},
      archivePrefix={arXiv},
      primaryClass={cs.CL},
      url={https://arxiv.org/abs/2402.14848}, 
}

@misc{neo4j,
  author       = {Neo4j, Inc.},
  title        = {Neo4j Graph Database},
  year         = {2025},
  howpublished = {\url{https://neo4j.com}}
}

@inproceedings{cognitive_debugging_1,
  title={Cognitive process during program debugging},
  author={Xu, Shaochun and Rajlich, V{\'a}clav},
  booktitle={Proceedings of the Third IEEE International Conference on Cognitive Informatics, 2004.},
  pages={176--182},
  year={2004},
  organization={IEEE}
}

@inproceedings{cognitive_debugging_2,
  title={Cognitive activities and support in debugging},
  author={Yoon, Byung-do and Garcia, Oscar N},
  booktitle={Proceedings fourth annual symposium on human interaction with complex systems},
  pages={160--169},
  year={1998},
  organization={IEEE}
}

@article{cognitive_debugging_3_icse,
  title={Towards a Cognitive Model of Dynamic Debugging: Does Identifier Construction Matter?},
  author={Hu, Danniell and Santiesteban, Priscila and Endres, Madeline and Weimer, Westley},
  journal={IEEE Transactions on Software Engineering},
  year={2024},
  publisher={IEEE}
}

@inproceedings{prompt_best_practice,
author = {White, Jules and Fu, Quchen and Hays, Sam and Sandborn, Michael and Olea, Carlos and Gilbert, Henry and Elnashar, Ashraf and Spencer-Smith, Jesse and Schmidt, Douglas C.},
title = {A Prompt Pattern Catalog to Enhance Prompt Engineering with ChatGPT},
year = {2023},
isbn = {9781941652190},
publisher = {The Hillside Group},
address = {USA},
booktitle = {Proceedings of the 30th Conference on Pattern Languages of Programs},
articleno = {5},
numpages = {31},
location = {Monticello, IL, USA},
series = {PLoP '23}
}

@article{CoT,
  title={Chain-of-thought prompting elicits reasoning in large language models},
  author={Wei, Jason and Wang, Xuezhi and Schuurmans, Dale and Bosma, Maarten and Xia, Fei and Chi, Ed and Le, Quoc V and Zhou, Denny and others},
  journal={Advances in neural information processing systems},
  volume={35},
  pages={24824--24837},
  year={2022}
}

@article{fewshot_CoT,
  title={Prompting large language models with chain-of-thought for few-shot knowledge base question generation},
  author={Liang, Yuanyuan and Wang, Jianing and Zhu, Hanlun and Wang, Lei and Qian, Weining and Lan, Yunshi},
  journal={arXiv preprint arXiv:2310.08395},
  year={2023}
}

@inproceedings{meta-prompting,
  title={Large language models are human-level prompt engineers},
  author={Zhou, Yongchao and Muresanu, Andrei Ioan and Han, Ziwen and Paster, Keiran and Pitis, Silviu and Chan, Harris and Ba, Jimmy},
  booktitle={The eleventh international conference on learning representations},
  year={2022}
}

@INPROCEEDINGS{bug_vs_feature,
  author={Herzig, Kim and Just, Sascha and Zeller, Andreas},
  booktitle={2013 35th International Conference on Software Engineering (ICSE)}, 
  title={It's not a bug, it's a feature: How misclassification impacts bug prediction}, 
  year={2013},
  volume={},
  number={},
  pages={392-401},
  doi={10.1109/ICSE.2013.6606585}}

@misc{qwen3coder2024,
  title={Qwen3-Coder: Large Mixture-of-Experts Model for Agentic Coding},
  author={Alibaba Cloud},
  howpublished={\url{https://github.com/QwenLM/Qwen3-Coder}}
}

@misc{devstral2024,
  title={DevStral: A New State-of-the-Art Open Model for Coding Agents},
  author={All Hands AI and Mistral AI},
  howpublished={\url{https://www.all-hands.dev/blog/devstral-a-new-state-of-the-art-open-model-for-coding-agents}}
}

@article{llama3herd2024,
  title={The Llama 3 Herd of Models},
  author={Meta AI},
  url={https://arxiv.org/abs/2407.21783}
}

@misc{langgraph2024,
  title={LangGraph: A Library for Building Stateful Multi-Agent Workflows},
  author={LangChain},
  howpublished={\url{https://github.com/langchain-ai/langgraph}}
}

@article{cdnn_2023,
title = {Automated software bug localization enabled by meta-heuristic-based convolutional neural network and improved deep neural network},
journal = {Expert Systems with Applications},
volume = {232},
pages = {120562},
year = {2023},
issn = {0957-4174},
doi = {https://doi.org/10.1016/j.eswa.2023.120562},
author = {Waqas Ali and Lili Bo and Xiaobing Sun and Xiaoxue Wu and Saifullah Memon and Saima Siraj and Ann {Suwaree Ashton}}
}

@article{blaze,
   title={BLAZE: Cross-Language and Cross-Project Bug Localization via Dynamic Chunking and Hard Example Learning},
   volume={51},
   ISSN={2326-3881},
   url={http://dx.doi.org/10.1109/TSE.2025.3579574},
   DOI={10.1109/tse.2025.3579574},
   number={8},
   journal={IEEE Transactions on Software Engineering},
   publisher={Institute of Electrical and Electronics Engineers (IEEE)},
   author={Chakraborty, Partha and Alfadel, Mahmoud and Nagappan, Meiyappan},
   year={2025},
   month=aug, pages={2254–2267} }

@misc{samir2025improvingirbasedbuglocalization,
      title={Improving IR-based Bug Localization with Semantics-Driven Query Reduction}, 
      author={Asif Mohammed Samir and Mohammad Masudur Rahman},
      year={2025},
      eprint={2510.04468},
      archivePrefix={arXiv},
      primaryClass={cs.SE},
      url={https://arxiv.org/abs/2510.04468}, 
}

@inproceedings{ReAct,
  title={React: Synergizing reasoning and acting in language models},
  author={Yao, Shunyu and Zhao, Jeffrey and Yu, Dian and Du, Nan and Shafran, Izhak and Narasimhan, Karthik R and Cao, Yuan},
  booktitle={The eleventh international conference on learning representations},
  year={2022}
}

@misc{scratchpad_agent_langchain,
  author    = {{LangChain}},
  title     = {Agent Scratchpad},
  url = {https://langchain-ai.github.io/langgraph/concepts/multi\_agent/}
}

@book{cormen_algorithms,
  author    = {Thomas H. Cormen and Charles E. Leiserson and Ronald L. Rivest and Clifford Stein},
  title     = {Introduction to Algorithms},
  edition   = {3rd},
  publisher = {MIT Press},
  address   = {Cambridge, MA},
  year      = {2009},
  isbn      = {978-0-262-03384-8}
}

@inproceedings{temperature_effect,
  title={The effect of sampling temperature on problem solving in large language models},
  author={Renze, Matthew},
  booktitle={Findings of the association for computational linguistics: EMNLP 2024},
  pages={7346--7356},
  year={2024}
}

@article{context_engineering,
  title={A survey of context engineering for large language models},
  author={Mei, Lingrui and Yao, Jiayu and Ge, Yuyao and Wang, Yiwei and Bi, Baolong and Cai, Yujun and Liu, Jiazhi and Li, Mingyu and Li, Zhong-Zhi and Zhang, Duzhen and others},
  journal={arXiv preprint arXiv:2507.13334},
  year={2025}
}

@article{context_irrelevant_confuse_llm,
  title={How easily do irrelevant inputs skew the responses of large language models?},
  author={Wu, Siye and Xie, Jian and Chen, Jiangjie and Zhu, Tinghui and Zhang, Kai and Xiao, Yanghua},
  journal={arXiv preprint arXiv:2404.03302},
  year={2024}
}

@misc{program_semantics,
      title={SemCoder: Training Code Language Models with Comprehensive Semantics Reasoning}, 
      author={Yangruibo Ding and Jinjun Peng and Marcus J. Min and Gail Kaiser and Junfeng Yang and Baishakhi Ray},
      year={2024},
      eprint={2406.01006},
      archivePrefix={arXiv},
      primaryClass={cs.CL},
      url={https://arxiv.org/abs/2406.01006}, 
}

@inproceedings{Grace_graph_bl,
  author    = {Lou, Yiling and Zhu, Qihao and Dong, Jinhao and Li, Xia and Sun, Zeyu and Hao, Dan and Zhang, Lu and Zhang, Lingming},
  title     = {Boosting Coverage-Based Fault Localization via Graph-Based Representation Learning},
  booktitle = {ESEC/FSE},
  year      = {2021},
  doi       = {10.1145/3468264.3468580}
}

@inproceedings{sgAttention_graph_bl,
  author    = {Tian, Yijun and Mei, Lijun and Yu, Yue and Wang, Shaowei and Wang, Haoyu and Liu, Yang},
  title     = {Structural-Guided Attention for Bug Localization},
  booktitle = {IJCAI},
  year      = {2023},
  doi       = {10.24963/ijcai.2023/461}
}

@misc{agentless_replication,
  title        = {Agentless Replication Package},
  author       = {{OpenAutoCoder}},
  year         = {2024},
  howpublished = {\url{https://github.com/OpenAutoCoder/Agentless}}
}

@misc{locagent_replication,
  title        = {LocAgent Replication Package},
  author       = {{Gerstein Lab}},
  howpublished = {\url{https://github.com/gersteinlab/LocAgent}}
}

@misc{langgraph_state,
  author       = {{LangChain, Inc.}},
  title        = {Low-level Concepts -- {\itshape LangGraph}},
  howpublished = {\url{https://langchain-ai.github.io/langgraph/concepts/low_level/}}
}

@misc{cognigent_replication,
	title        = {CogniGent: Replication Package},
	howpublished = {\url{https://github.com/asifsamir/CogniGent/}},
}

@online{ctrl_click,
  author       = {Augustin Popa},
  title        = {Productivity Improvements: Ctrl + Click Go to Definition},
  year         = {2017},
  month        = oct,
  day          = {11},
  url          = {https://devblogs.microsoft.com/cppblog/productivity-structure-visualizer-ctrl-click-to-go-to-definition/},
  note         = {Microsoft Dev Blogs}
}

@misc{openai_gpt4,
  title        = {GPT-4 Technical Report},
  author       = {{OpenAI}},
  year         = {2023},
  eprint       = {2303.08774},
  archivePrefix= {arXiv},
  primaryClass = {cs.CL},
  url          = {https://arxiv.org/abs/2303.08774}
}

@inproceedings{stochasticity_LLM,
    title = "The Good, The Bad, and The Greedy: Evaluation of {LLM}s Should Not Ignore Non-Determinism",
    author = "Song, Yifan  and
      Wang, Guoyin  and
      Li, Sujian  and
      Lin, Bill Yuchen",
    editor = "Chiruzzo, Luis  and
      Ritter, Alan  and
      Wang, Lu",
    booktitle = "Proceedings of the 2025 Conference of the Nations of the Americas Chapter of the Association for Computational Linguistics: Human Language Technologies (Volume 1: Long Papers)",
    month = apr,
    year = "2025",
    address = "Albuquerque, New Mexico",
    publisher = "Association for Computational Linguistics",
    url = "https://aclanthology.org/2025.naacl-long.211/",
    doi = "10.18653/v1/2025.naacl-long.211",
    pages = "4195--4206",
    ISBN = "979-8-89176-189-6"
}

\appendix









\end{document}